\documentclass[12pt,a4paper,twoside]{article}

\usepackage[applemac]{inputenc}
\usepackage[english]{babel}
\usepackage[T1]{fontenc}
\usepackage{lmodern}
\usepackage{geometry} 
\usepackage[intlimits,sumlimits]{amsmath}
\usepackage{amssymb}
\usepackage[amsthm,thmmarks]{ntheorem}
\usepackage{mathrsfs}
\usepackage{bm}
\usepackage{xpatch}
\usepackage{dsfont}
\usepackage{siunitx} 
\usepackage{scalerel,stackengine}
\usepackage{aligned-overset}
\usepackage{algorithm}
\usepackage{multicol}
\usepackage{booktabs}
\usepackage{makecell}

\usepackage[backend=bibtex, style=authoryear, firstinits=true, maxcitenames=2, sorting=nty, isbn=false, dashed=false]{biblatex}
\DeclareNameAlias{sortname}{last-first}
\DeclareFieldFormat*{title}{#1}
\DeclareFieldFormat*{edition}{#1 edition}
\DeclareDelimFormat[cite,parencite]{nameyeardelim}{\textcolor{black}{\addcomma\space}}

\DeclareFieldFormat{citehyperref}{%
   \DeclareFieldAlias{bibhyperref}{noformat}
   \bibhyperref{#1}}

\savebibmacro{cite}

\renewbibmacro*{cite}{%
   \printtext[citehyperref]{%
      \restorebibmacro{cite}%
      \usebibmacro{cite}}}
\usepackage{xpatch}
\xpatchbibmacro{textcite}
  {\printnames{labelname}}
  {\printtext[bibhyperref]{\printnames{labelname}}}
  {}{}

\usepackage{caption}
\usepackage{subcaption}
\usepackage{paralist}
\usepackage{graphicx}
\usepackage{colortbl}
\usepackage[percent]{overpic}
\usepackage{mathtools}
\usepackage[usenames,dvipsnames,svgnames,table]{xcolor}
\usepackage{pgf,tikz}
\usetikzlibrary{patterns}
\usetikzlibrary{quotes,babel,angles}
\usetikzlibrary{arrows}
\usepackage{rotating}
\usepackage{pgfplots}
\usepackage{wrapfig}
\usepackage[most]{tcolorbox}
\usepackage[compact]{titlesec}
\usepackage{fancyhdr}

\definecolor{UHwhite}{RGB}{255,255,255}
\definecolor{UHorange}{RGB}{255,128,0}
\definecolor{UHred}{RGB}{202,53,56}
\definecolor{UHgruen}{RGB}{161,176,45}
\definecolor{UHdarkblue}{RGB}{6,68,107}
\definecolor{UHblue}{RGB}{24,105,183}
\definecolor{UHmidblue}{RGB}{17,119,182}
\definecolor{UHcyan}{RGB}{87,168,211}
\definecolor{UHbrightblue}{RGB}{160,206,234}
\definecolor{UHdarkgrey}{RGB}{69,69,69}
\definecolor{UHgrey}{RGB}{148,148,148}
\definecolor{UHbrightgrey}{RGB}{215,215,215}
\definecolor{UHbeige}{RGB}{235,230,215}

\definecolor{myred}{RGB}{178,34,34}
\definecolor{mygrey}{RGB}{69,69,69}

\numberwithin{equation}{section}
\definecolor{blau}{RGB}{24,105,183}
\usepackage[colorlinks=False,
pdfpagelabels,
pdfstartview = FitH,
bookmarksopen = true,
bookmarksnumbered = true,
colorlinks   = true,
linkcolor = black,
citecolor = UHblue,
plainpages = false,
hypertexnames = false,
linkbordercolor = UHblue,
urlcolor     = UHblue,
citebordercolor=white,
urlbordercolor=white]{hyperref}
\usepackage{orcidlink}
\usepackage{sectsty}
\usepackage{booktabs}

\usepackage[justification=centering]{caption}
\usepackage{dcolumn}
\usepackage{framed}
\newcolumntype{2}{D{.}{}{3.0}}
\makeatletter 
\renewcommand*{\@pnumwidth}{3em}      
\makeatother
\makeatletter
\renewcommand{\p@envcount}{\thesubsection.}
\makeatother

\newcommand{\R}{\mathbb{R}}

\newcommand{\mng}[1]{\left \{ #1 \right \}}

\newcommand{\spn}[1]{\mathrm{span}\kern-0.4ex\left(#1\right)}
\newcommand{\supp}[1]{\mathrm{supp}\kern-0.4ex\left(#1\right)}
\newcommand{\bild}[1]{\mathrm{Bild}\kern-0.4ex\left( #1\right)}
\newcommand{\krn}[1]{\mathrm{Kern}\kern-0.4ex\left( #1\right)}
\newcommand{\rang}[1]{\mathrm{rang}\kern-0.4ex\left( #1\right)}
\newcommand{\spur}[1]{\mathrm{Spur}\kern-0.4ex\left( #1\right)}

\newcommand{\ggt}[1]{\mathrm{ggT}\kern-0.4ex\left( #1\right)}
\newcommand{\kgv}[1]{\mathrm{kgV}\kern-0.4ex\left( #1\right)}

\newcommand{\grpGL}[1]{\mathrm{GL}\kern-0.4ex\left( #1\right)}
\newcommand{\grpSL}[1]{\mathrm{SL}\kern-0.4ex\left( #1\right)}
\newcommand{\grpO}[1]{\mathrm{O}\kern-0.4ex\left( #1\right)}
\newcommand{\grpSO}[1]{\mathrm{SO}\kern-0.4ex\left( #1\right)}
\newcommand{\grpU}[1]{\mathrm{U}\kern-0.4ex\left( #1\right)}
\newcommand{\grpSU}[1]{\mathrm{SU}\kern-0.4ex\left( #1\right)}

\renewcommand{\P}[1]{\mathbb{P}\kern-0.4ex\left( #1\right)}
\newcommand{\E}[1]{\mathbb{E}\kern-0.4ex\left( #1\right)}

\makeatletter
\renewenvironment{abstract}{%
    \if@twocolumn
      \section*{\abstractname}%
    \else 
      \begin{center}%
        {\bfseries \abstractname\vspace{\z@}}%
      \end{center}%
      \quotation
    \fi}
    {\if@twocolumn\else\endquotation\fi}
\makeatother

\usepackage{titling}
\begin{document}
\sloppy

\title{\fontsize{24}{30} \selectfont \textbf{D-Vine GAM Copula based Quantile Regression with Application to Ensemble Postprocessing}}

\author{David Jobst\,\orcidlink{0000-0002-2014-3530}\thanks{Corresponding author, University of Hildesheim, Institute of Mathematics and Applied Informatics, Samelsonplatz 1, 31141 Hildesheim, Germany, \texttt{\href{mailto:jobstd@uni-hildesheim.de}{jobstd@uni-hildesheim.de}}},\and  Annette M\"oller\,\orcidlink{0000-0001-9386-1691}\thanks{Bielefeld University, Faculty of Business Administration and Economics, Universit\"atsstra{\ss}e 25, 33615 Bielefeld, Germany, \texttt{\href{mailto:annette.moeller@uni-bielefeld.de}{annette.moeller@uni-bielefeld.de}}} \thanks{Helmholtz Centre for Infection Research (HZI), Inhoffenstra{\ss}e 7, 38124 Braunschweig, Germany, \texttt{\href{mailto:annette.moeller@helmholtz-hzi.de}{annette.moeller@helmholtz-hzi.de}}},\and J\"urgen Gro{\ss}\,\orcidlink{0000-0002-3861-4708}\thanks{University of Hildesheim, Institute of Mathematics and Applied Informatics, Samelsonplatz 1, 31141 Hildesheim, Germany, \texttt{\href{mailto:juergen.gross@uni-hildesheim.de}{juergen.gross@uni-hildesheim.de}}}}
\maketitle
\thispagestyle{empty}

\begin{abstract}
\small Temporal, spatial or spatio-temporal probabilistic models are frequently used for weather forecasting. The D-vine (drawable vine) copula quantile regression (DVQR) is a powerful tool for this application field, as it can automatically select important predictor variables from a large set and is able to model complex nonlinear relationships among them. However, the current DVQR does not always explicitly and economically allow to account for additional covariate effects, e.g.  temporal or spatio-temporal information. Consequently, we propose an extension of the current DVQR, where we parametrize the bivariate copulas in the D-vine copula through Kendall's $\tau$ which can be linked to additional covariates. The parametrization of the correlation parameter allows generalized additive models (GAMs) and spline smoothing to detect potentially hidden covariate effects. The new method is called GAM-DVQR, and its performance is illustrated in a case study for the postprocessing of 2\,\si{\meter} surface temperature forecasts. We investigate a constant as well as a time-dependent Kendall's $\tau$. The GAM-DVQR models are compared to the benchmark methods Ensemble Model Output Statistics (EMOS), its gradient-boosted extension (EMOS-GB) and basic DVQR. The results indicate that the GAM-DVQR models are able to identify time-dependent correlations as well as relevant predictor variables and significantly outperform the state-of-the-art methods EMOS and EMOS-GB. Furthermore, the introduced parameterization allows using a static training period for GAM-DVQR, yielding a more sustainable model estimation in comparison to DVQR using a sliding training window. Finally, we give an outlook of further applications and extensions of the GAM-DVQR model. To complement this article, our method is accompanied by an \texttt{R}-package called \texttt{gamvinereg} on \href{https://github.com/jobstdavid/gamvinereg}{GitHub}. 

\end{abstract}
\textbf{Keywords:} conditional copula; vine copula; dependence modeling; quantile regression; covariate effects; ensemble postprocessing; probabilistic forecasting.

\newpage

\section{Introduction}

Nowadays, weather forecasts are based on numerical weather prediction (NWP) models which suffer from various uncertainties. In practice, ensemble prediction systems (EPS) are commonly used to address these uncertainties. Therefore, the NWP model is run multiple times with different model and/or initial and boundary conditions \parencite{Gneiting2005, Leutbecher2008}. Afterwards a forecast ensemble is generated, which can be seen as a probabilistic forecast allowing to quantify forecast uncertainty \parencite{Palmer2002}.

However, the forecast ensemble usually suffers from biases and dispersion errors and thus may benefit from statistical postprocessing using past data to improve calibration and forecast skill. A popular postprocessing model is the so called Ensemble Model Output Statistics (EMOS, \cite{Gneiting2005}). This method is used to obtain a full predictive distribution from the ensemble forecasts. Originally, EMOS was developed for Gaussian distributed weather quantities, e.g. temperature or air pressure. Later, machine learning methods such as quantile regression forests (QRF, \cite{Taillardat2016}) or gradient boosting EMOS (EMOS-GB, \cite{Messner2017}) have been investigated to extend the classical EMOS setting. \textcite{Rasp2018} compared distributional regression networks to QRF as well as EMOS-GB for the postprocessing of  2\,\si{\meter} surface temperature forecasts and found only minor differences among these methods for longer training periods. Recently, the D-vine copula based quantile regression (DVQR), which was developed by \textcite{Kraus2017} and further extended by \textcites{Tepegjozova2022, Sahin2022}, was used by \textcites{Moeller2018, Demaeyer2023} for the postprocessing of 2\,\si{\meter} surface temperature forecasts. \textcite{Jobst2023} used the same method for the postprocessing of 10\,\si{\meter} surface wind speed forecasts. In all three analyses, DVQR showed comparable or sometimes even better results with respect to its competing methods. 

Reasons for the superior performance of DVQR are manifold. DVQR  is a quantile regression model that overcomes typical issues of quantile regression such as quantile crossings, transformations, collinearity and the integration of interactions of variables \parencite{Kraus2017}. In addition, DVQR is able to model complex nonlinear relationships between the explanatory variables and response while imposing less restrictive model assumptions. Last but not least, it can theoretically adapt any distribution shape. 

One drawback in the current DVQR is the fact that it is not straightforward to explicitly include covariate effects, such as temporal effects into the model. This is one reason for estimating DVQR by sliding training windows, where the complete model needs to be re-estimated for each prediction time point. This can become computationally expensive, as the optimal sliding window size is not known in advance and additionally needs to be determined. In the ensemble postprocessing context the sliding window size depends on various factors, such as considered variables, seasons, locations, etc., which should be ideally taken into account. Therefore, \textcite{Jobst2023} compared different types of training periods in the DVQR model estimation, and detected that a reduction in the computational complexity usually comes along with a worse predictive performance. 

In this work, we exactly tackle this problem and allow for arbitrary covariate effects in the DVQR model, such as temporal, spatial or spatio-temporal ones. To be more precise, the correlation among two variables according to Kendall's $\tau$ can depend on covariates and is subsequently used to calculate the parameters for the bivariate copulas in the D-vine copula. For this, we combine the work of \textcites{Vatter2015, Vatter2018} who introduced parametric bivariate copulas and later vine copulas depending on covariates with the DVQR proposed by \textcite{Kraus2017}. As the correlations linked to the copulas are parametrized in terms of generalized additive models (GAMs, \cite{Hastie1990, Green1993}) and smoothing splines our method will be called GAM-DVQR.

We apply the GAM-DVQR with covariates modeling temporal effect for the postprocessing of 2\,\si{\meter} surface temperature forecasts at 462 observation stations in Germany. The results show that our suggested method is able to capture temporal covariate effects and can select important predictor variables from a potentially large set. Furthermore, the correlation time-dependent GAM-DVQR models show better results in comparison to the GAM-DVQR model assuming constant correlations, and are able to significantly outperform the benchmark methods EMOS and EMOS-GB. Last but not least, due to the use of a static training period for GAM-DVQR, the model needs to be fitted only once which is more efficient than estimating DVQR on a sliding window and therefore makes it attractive for practical and operational use. To the best of the authors knowledge, GAM-DVQR has not been suggested yet and further analyzed in an application. Although the presented application of GAM-DVQR is concerned with the meteorological field, our suggested method is broadly applicable to various areas where any correlation dependent covariate effects are required to be integrated. 

The rest of the paper is organized as follows: Section \ref{sec: D-vine copula based quantile regression methods} introduces the D-vine copula based quantile regression methods including DVQR and GAM-DVQR. In Section \ref{sec: Data}, the data set for our application is described. A brief overview of the competing ensemble postprocessing methods is given in Section \ref{sec: Methods}. Section \ref{sec: Verification} provides a short introduction to the commonly used verification measures in the ensemble postprocessing field. In Section \ref{sec: Results}, we discuss the results of our application. We close with a conclusion and outlook in Section \ref{sec: Conclusion and outlook}.  

\section{D-vine copula based quantile regression methods}
\label{sec: D-vine copula based quantile regression methods}

In this section, we outline the copula method requirements 
employed by our postprocessing approaches.

\subsection{Copulas}
Multivariate standard distributions, e.g. the multivariate normal are often restricted in their marginal behavior, as they assume that all marginals are of the same type. The application of copulas allows to overcome this problem. A $p$-dimensional \textit{copula} $C$ is a multivariate distribution function on $[0,1]^p$. According to Sklar's theorem \parencite{Sklar1959}, for every multivariate distribution function $F$ of $p$ continuous variables $\bm{X}:=(X_1,\ldots, X_p)\in \R^p$ there exists a copula $C$, such that 
\begin{align}
	F(x_1,\ldots, x_p)=C(F_1(x_1),\ldots, F_p(x_p)),
\end{align}
where $F_j$ for $j=1,\ldots, p$ are the marginal distribution functions and $\bm{x}:=(x_1,\ldots,x_p)\in \R^p$ are the realizations of $\bm{X}$. 
If all distribution functions are differentiable, the corresponding $p$-dimensional joint density function $f$ can be expressed by
\begin{align}
	f(x_1,\ldots, x_p)=c(F_1(x_1),\ldots, F_p(x_p))\cdot f_1(x_1)\cdots f_p(x_p),
\end{align}
where $c$ denotes the copula density function of the copula $C$ and $f_j$ for $j=1,\ldots, p$ are the marginal density functions of the variables $X_1,\ldots, X_p$.

Therefore, every multivariate distribution function $F$ can be decomposed into a copula $C$ modeling the dependence and its univariate marginal distributions allowing to construct a wide range of distributions.

\subsection{D-vine copula}
Multivariate copulas such as e.g. the elliptical and the archimedean copulas are often not adaptable enough, as they usually presume that the variables in all the pairs have homogeneous dependence structures. \textcite{Bedford2001, Bedford2002} extended the theory about multivariate copulas by developing the so-called pair-copula construction (PCC), where the joint dependence is build up by only bivariate copulas using conditioning. As a PCC is not unique, \textcite{Bedford2002} introduced a graphical structure which is called \textit{regular vine}. A regular vine consists of a set of nested trees,  where the edges in one tree become the nodes of the subsequent one. In a regular vine consisting of $p$ variables, the nodes and edges in the first tree represent the $p$ variables and unconditional dependence for $p-1$ variables, respectively. In the subsequent trees the conditional dependence of a pair of variables conditioned on the variables they have in common is modeled. A \textit{regular vine copula} is obtained by specifying bivariate copulas, so called pair-copulas, on each edge of the trees.

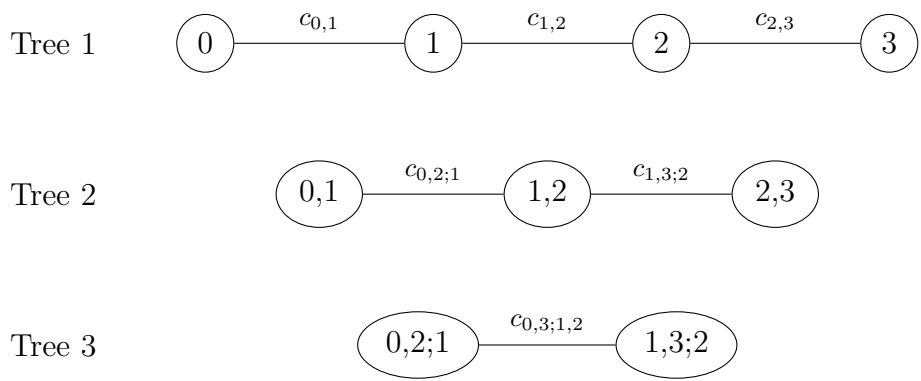
\begin{figure}[h!]
\begin{center}
\begin{tikzpicture}[scale = 2]
\usetikzlibrary{shapes}
	\node (Tree 1) at (-1, 0) {Tree 1};
	\node (Tree 2) at (-1, -1) {Tree 2};
	\node (Tree 3) at (-1, -2) {Tree 3};

      \node[ellipse, minimum size=0.75cm, draw=black] (1) at (0,0) {0};
      \node[ellipse, minimum size=0.75cm, draw=black] (2) at (1.5, 0) {1};
      \node[ellipse, minimum size=0.75cm, draw=black] (3) at (3, 0) {2};
      \node[ellipse, minimum size=0.75cm, draw=black] (4) at (4.5,0) {3};
			
\draw (1)--node[midway,above,sloped]{\footnotesize $c_{0,1}$}(2)
			(2)--node[midway,above,sloped]{\footnotesize $c_{1,2}$}(3)
			(3)--node[midway,above,sloped]{\footnotesize $c_{2,3}$}(4);			

      \node[ellipse, minimum size=0.75cm, draw=black] (1) at (0.75,-1) {0,1};
      \node[ellipse, minimum size=0.75cm, draw=black] (2) at (2.25, -1) {1,2};
      \node[ellipse, minimum size=0.75cm, draw=black] (3) at (3.75, -1) {2,3};
			
\draw (1)--node[midway,above,sloped]{\footnotesize $c_{0,2;1}$}(2)
			(2)--node[midway,above,sloped]{\footnotesize $c_{1,3;2}$}(3);	

      \node[ellipse, minimum size=0.75cm, draw=black] (1) at (1.4,-2) {0,2;1};
      \node[ellipse, minimum size=0.75cm, draw=black] (2) at (3.1, -2) {1,3;2};
			
\draw (1)--node[midway,above,sloped]{\footnotesize $c_{0,3;1,2}$}(2);			
			
\end{tikzpicture}	
\end{center}
	\caption{4-dimensional D-vine tree structure and corresponding pair-copula densities.}\label{fig: D-vine}
\end{figure}

A \textit{D-vine} is special class of a regular vine in which each tree is a path, i.e. all nodes in the graph are connected to at most two others (see Figure \ref{fig: D-vine}). 
Therefore, a \textit{D-vine copula} is a regular vine copula, where the tree structure is a D-vine. A node in a D-vine copula represents a certain variable, while an edge between a pair of nodes corresponds to the dependence among the variables associated with the respective nodes expressed by a pair-copula. 

For a short overview about vine copulas see e.g. \textcite{Czado2022} and for a detailed introduction see e.g. \textcite{Czado2019}.   

\subsection{D-vine copula based quantile regression}

D-vine copulas can be used in a univariate or multivariate regression context. Our focus will be on the univariate setting, where it is possible to derive a conditional D-vine copula density. For this, the leaf node in the first tree of a D-vine needs to be the response variable. In the following, we denote $Y$ as response variable with marginal distribution function  $F_Y$ and the $p$ predictor variables by $X_1,\ldots,X_p$ with marginal distribution functions $F_1,\ldots, F_p$. The lower case letters of the response and predictor variables represent the respective realizations. For a D-vine copula with node order $(0, 1,\ldots, p)$ corresponding to the variable order $(Y,X_1,\ldots,X_p)$, $p\geq 2$, the conditional density of $Y$ given $X_1,\ldots, X_p$ can be obtained by
\begin{align}
f_{0\vert 1,\ldots, p}(y\vert x_1,\ldots,x_p)&=\prod\limits_{j=2}^{p}c_{0,j;1,\ldots, j-1}(F_{0\vert 1,\ldots, j-1}(y\vert x_1,\ldots,x_{j-1})\notag,
							  F_{j\vert 1,\ldots, j-1}(x_j\vert x_1,\ldots, x_{j-1}))\\
							  &\qquad\qquad \cdot c_{0,1}(F_Y(y),F_1(x_1))\cdot f_Y(y),
\end{align}
where $F_{0\vert 1,\ldots, j-1}$ and $F_{j\vert 1,\ldots, j-1}$ denote the  distribution functions of the conditional random variables $Y\vert X_1=x_1,\ldots, X_{j-1}=x_{j-1}$ and $X_j = x_j \vert X_1 = x_1, \ldots, X_{j-1}=x_{j-1}$, respectively, and can be calculated recursively. Furthermore, $c_{0, j;1,\ldots, j-1}$ denotes the bivariate copula (pair-copula) density of the bivariate distribution of $(Y, X_j)$ given $X_{1}=x_{1}, \ldots, X_{j-1}=x_{j-1}$. To allow for easy estimation, we make the simplifying assumption \parencite{Stoeber2013}, i.e. we assume that the pair-copulas of conditional distributions are independent of the values of variables on which they are conditioned. Nevertheless, the pair-copula densities of the higher tree levels depend on the conditioning values by its arguments. 

Based on the conditional D-vine copula distribution, \textcite{Kraus2017} introduced a quantile regression (DVQR). The conditional quantile function for a D-vine copula with $p$ predictor variables $X_1,\ldots,X_p$ at quantile level $\alpha\in (0,1)$ can be calculated as 

\begin{align}
F^{-1}_{0\vert 1,\dots, p}(\alpha\vert x_1,\dots, x_p):=F_Y^{-1}\left(C_{0\vert 1, \ldots, p}^{-1}(\alpha\vert F_1(x_1),\dots, F_p(x_p))\right),
\label{eq: cond. d-vine copula}
\end{align}

where $F_Y^{-1}$ is the inverse marginal distribution of the response variable $Y$ and $C_{0\vert 1, \ldots, p}^{-1}$ denotes the conditional D-vine copula quantile function. 

\paragraph{Estimation Procedure.}

The estimation of the D-vine copula quantile regression follows a two-step procedure called ``inference for margins'' \parencite{Joe1996b}. Firstly, the marginal distributions of all variables are estimated. This is necessary for transforming the raw data of each variable to uniformly distributed data in $[0,1]$ by the probability integral transformation (PIT). Consequently, we obtain the realizations $v=F_Y(y)$ and  $u_i=F_i(x_i)$ of the random variables $V=F_Y(Y)$ and $U_i=F_i(X_i)$ for $i=1,\ldots, p$. The marginal distributions can be estimated parametrically or non-parametrically. Secondly, the conditional copula function $C_{0\vert 1, \ldots, p}$ can be obtained in a closed form by a composition of so-called $h$-functions associated with the pair-copulas \parencite{Joe1996a}. This two-step approach is very often preferred over estimating the marginal distributions and copulas simultaneously, as the joint estimation may be harder to implement, can be very time consuming or is sometimes simply infeasible.

\paragraph{Order of Variables.}
The only remaining question is in which order the PIT transformed variables $V,U_1,\ldots,U_p$ should appear in the D-vine. By construction, the transformed response variable $V$ needs to be the leaf node in the first tree of the D-vine (node 0 in Figure \ref{fig: D-vine}). As the order of the predictors $U_1,\ldots, U_p$ is usually not obviously predetermined, one can select the most informative predictors and order them according to their predictive strength. To do so, \textcite{Kraus2017} propose a sequential forward selection approach to select the most important predictors by improving an (AIC/BIC)-corrected conditional log-likelihood. 

For the demonstration of the DVQR algorithm, we assume, that $k-1$ predictors have already been selected and the current D-vine has the ordering $(V, U_{l_1},\ldots,U_{l_{k-1}})$, where $\mng{l_1,\ldots, l_{k-1}}\subset \mng{1,\ldots, p}$. Using each remaining predictor $U_j$ with $j\in \mng{1,\ldots, p}\backslash \mng{l_1,\ldots, l_{k-1}}$, the current D-vine is estimated for $(V,U_{l_1},\ldots,U_{l_{k-1}},U_j)$. In each step of the DVQR method, the optimal pair-copulas according to the minimum AIC/BIC-conditional log-likelihood or maximum conditional log-likelihood are chosen. Having estimated the necessary pair-copulas to extend the current D-vine for each of the $U_j$, we update the model by adding the variable which yields to the lowest AIC/BIC- or highest conditional log-likelihood of the model.

\subsection{D-vine GAM copula based quantile regression}
\label{sec: D-vine GAM copula based quantile regression}

In the original formulation of DVQR, \textcite{Kraus2017} assume parametric pair-copulas, where the copula parameters are constant. A natural extension of the parametric pair-copulas includes additional effects of covariates, e.g. in time and/or space into the copula parameters by modelling them as functions of such covariates. This statistical tool is called \textit{conditional copula}, which has already been discussed earlier by \textcite{Patton2002} using a fully parametric approach for the copula parameter estimation. Later, \textcite{Gijbels2011} proposed a non-parametric version and \textcite{Acar2010} a semi-parametric conditional copula model. \textcite{Vatter2015} were the first to suggested an alternative approach based on generalized additive models (GAMs, \cite{Hastie1990, Green1993}) and spline smoothing for the copula parameter. While the previous mentioned approaches are restricted to bivariate copulas only, \textcite{Vatter2018} extend the idea of conditional copulas to higher dimensions. More precisely, they used the GAM based bivariate copula framework as suggested by \textcite{Vatter2015} for the construction of vine copulas. 

In our proposed D-vine GAM copula based quantile regression (GAM-DVQR), we use the GAM based bivariate copulas as suggested by \textcite{Vatter2015} for the D-vine copula quantile regression to include effects of covariates.  The sequential forward variable selection algorithm of DVQR is used in the same way for GAM-DVQR. The difference between these two methods is mainly in the estimation of the bivariate copulas, which will be briefly illustrated in the following. 

For a vector of $q$ covariates $\bm{Z}:=(Z_1,\ldots, Z_q)\in \R^q$ with realizations $\bm{z}:=(z_1,\ldots, z_q)\in \R^q$, a parametric form  is assumed for the conditional copula densities $c(\cdot,\ \cdot; \eta(\bm{z}))$, where the copula parameter $\eta(\bm{z})$ depends on the covariates $\bm{z}$. For frequently used copula families, bijective transformations between the copula parameter $\eta(\bm{z})$ and Kendall's $\tau(\bm{z})$ can be derived (see Table \ref{tab: par_tau_trafo}). Note, that additional exogenous variables which do not belong to the set of predictor variables in the D-vine such as, e.g. temporal, spatial or other variables can be chosen as covariates. Nonetheless, it is theoretically possible to incorporate predictor variables from the D-vine as covariates as well.

\begin{table}[h!]
\begin{center}
\begin{tabular}{c c c} 
\toprule
Copulas & $\eta(\bm{z})$ & $\tau(\bm{z})$ \\ \hline
Gaussian, Student-$t$ & $\sin\left(\frac{\pi}{2}\tau(\bm{z})\right)$ & $\frac{2}{\pi}\arcsin\left(\eta(\bm{z})\right)$ \\
Gumbel, Gumbel-180$^\circ$ & $\frac{1}{1-\tau(\bm{z})}$ & $1-\frac{1}{\eta(\bm{z})}$\\
Gumbel-90$^\circ$, Gumbel-270$^\circ$ & $-\frac{1}{1+\tau(\bm{z})}$ & $-1-\frac{1}{\eta(\bm{z})}$\\
Clayton, Clayton-180$^\circ$ & $\frac{2\tau(\bm{z})}{1-\tau(\bm{z})}$ & $\frac{\eta(\bm{z})}{\eta(\bm{z})+2}$\\
Clayton-90$^\circ$, Clayton-270$^\circ$ & $\frac{2\tau(\bm{z})}{1+\tau(\bm{z})}$ & $\frac{\eta(\bm{z})}{2-\eta(\bm{z})}$\\
\bottomrule
\end{tabular}
\end{center}
\caption{Mappings between the copula parameter and Kendall's $\tau$. The degrees represent the amount of rotation of the respective copula, e.g. a Gumbel-180$^\circ$ copula is a Gumbel copula rotated by 180$^\circ$ counterclockwise.}
\label{tab: par_tau_trafo}
\end{table}

Due to the one-to-one mappings between the copula parameters and Kendall's $\tau$, we re-parameterize all conditional copulas in the D-vine copula as functions of the corresponding Kendall's $\tau$ and write $c(\cdot,\ \cdot; \tau(\bm{z}))$. The modeling of the Kendall's $\tau$ instead of the actual copula parameter might seem unnecessary at first sight. However, two useful properties arise from this approach \parencite{Vatter2015}. Firstly, a dependence measure such as Kendall's $\tau$ is easier to interpret than a copula parameter. Secondly, this approach makes it simpler to compare different types of copula families, as there exists a natural relationship between the copula parameter and Kendall's $\tau$. If the actual copula parameter is modeled, it often becomes necessary to specify a link function to ensure the predefined range of the copula parameter. Depending on the copula family, different link functions need to be selected, resulting in possible miss-specifications of the link function (see, e.g., \cite{Li1989}) and in a comparison which is not standardized.

Therefore, \textcite{Vatter2015} suggest to model the change in the correlation with respect to the covariates as 

\begin{align}
	g^{-1}(\tau(\bm{u}, \bm{v};\bm{\alpha}, \bm{s})):=\bm{\alpha}\bm{u}^T+\sum\limits_{k=1}^{K}s_k(\bm{v}_k),
	\label{eq: tau_par}
\end{align}

where $g^{-1}(\tau):=2\cdot\mathrm{artanh}(\tau)$
is the inverse link function between the GAM and Kendall's $\tau$ to ensure the parameter range,
$\bm{u}\in \R^J$ and $\bm{v}\in \R^K$ are subsets of the covariate $\bm{z}$ or products thereof to consider interactions, and $\bm{\alpha}\in \R^J$ is a vector of parameters for the linear component. The mappings
$s_k:\mathbb{S}_k\to \mathbb{R}$ are smooth functions supported on closed intervals $\mathbb{S}_k\subset \R$ for $k=1,\ldots,K$, i.e. $s_k\in C^2(\mathbb{S}_k)$ admits a finite-dimensional basis-quadratic penalty representation such as natural cubic splines, cyclic cubic splines or tensor product splines. Moreover, $\bm{s}:=(\bm{s}_1,\ldots, \bm{s}_K)\in \R^M$ denotes the parameter vector for the $K$ smooth functions $s_k$ with a total of $M:=\sum_{k=1}^{K}m_k$ parameters.

Models as in Equation \eqref{eq: tau_par} are called partially linear models \parencite{Haerdle2000}, as they consist of a linear component $\bm{\alpha}\bm{u}^T$ and a non-linear component $\sum_{k=1}^{K}s_k(\bm{v}_k)$. The maximum penalized log-likelihood estimates of the parameters $\bm{\alpha},\bm{s}$ are obtained by iteratively reweighted generalized ridge regression. For a more technical description of the copula parameter estimation as well as for extensive simulation studies for the suggested conditional copulas and vine copulas, see \textcites{Vatter2015, Vatter2018}. Moreover, it should be mentioned that Kendall's $\tau$ will be modeled by only one single model specified in Equation \eqref{eq: tau_par} for all bivariate copulas in the D-vine copula. As stated above, we will use the term ``predictor variables'' for the variables in the D-vine and the term ``covariates'' for the variables included in Equation \eqref{eq: tau_par} for modeling the conditional bivariate copula.

To complement this work, \textcite{Jobst2023b} developed an \texttt{R}-package called \texttt{gamvinereg} for the GAM-DVQR, which is based on the R-package \texttt{gamCopula} by \textcite{Vatter2018} and on the code of the \texttt{R}-package \texttt{vinereg} by \textcite{Nagler2022}.

\section{Data}
\label{sec: Data}

\begin{figure}[h!]
	\begin{center}
		\includegraphics[scale = 0.55]{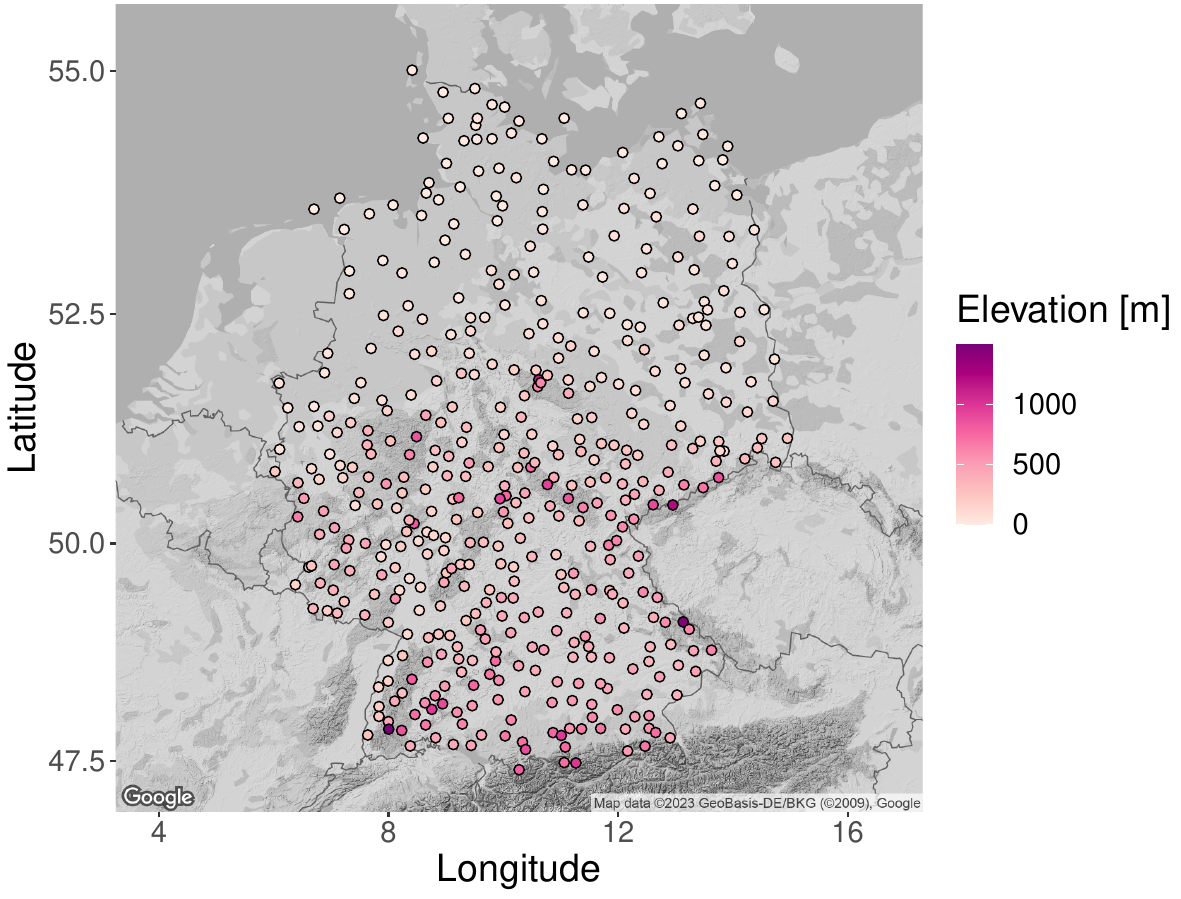}
	\end{center}	
	\caption{Observation stations for  2\,\si{\meter} surface temperature.}
	\label{fig: stations}
\end{figure}

To illustrate the capabilities of GAM-DVQR, we present a case study for the postprocessing of 2\,\si{\meter} surface temperature forecasts initialized at 1200 UTC for a forecast lead time of 24\,\si{\hour}. The 2\,\si{\meter} surface temperature observations are provided by \textcite{DWD2018} with maximal 5\% missing observations at each synoptic observation station between January 2, 2015 to December 31, 2020, which leads to 462 observation stations (see Figure \ref{fig: stations}). 

The ensemble forecasts are provided by the \textcite{ECMWF2021}, consisting of $m=50$ perturbed ensemble members. These forecasts are initialized at 1200 UTC on a grid of 0.25$^\circ$ $\times$ 0.25$^\circ$. The gridded data is bilinearly interpolated to the observation stations. 

Additionally to the target variable (2\,\si{\meter} surface temperature) we add several auxiliary ensemble predictor variables, for an overview see Table \ref{tab: pred_var}. We calculate the 10\,\si{\meter} surface wind speed by $\text{ws10m}:= \sqrt{\text{u10m} + \text{v10m}}$ and the 2\,\si{\meter} surface relative humidity is approximated by $\mathrm{r2m}:=\exp\left(\frac{17.625\cdot \mathrm{d2m}}{243.04+\mathrm{d2m}}\right)/\exp\left(\frac{17.625\cdot \mathrm{t2m}}{243.04+\mathrm{t2m}}\right)$ according to \textcite{Alduchov1996}. Furthermore, we calculate the variable sine- and cosine-transformed day of the year (doy) abbreviated as $\sin$ and $\cos$ via $\sin\left(\frac{2\pi\cdot \mathrm{doy}}{365.25}\right)$ and $\cos\left(\frac{2\pi\cdot \mathrm{doy}}{365.25}\right)$ for $\mathrm{doy}\in \mng{1,2,\ldots, 366}$, respectively.

In the following, ensemble forecasts are comprised to their mean and standard deviation, where for a weather variable $v$, 
\begin{equation}
\overline{X}_v:=\frac{1}{m}\sum\limits_{i=1}^{m}X_{v, i}\quad \text{and}\quad  S_v:=\sqrt{\frac{1}{m-1}\sum\limits_{i=1}^{m}(\overline{X}_v-X_{v, i})^2},
\end{equation}
will denote the ensemble mean and standard deviation, respectively, from an $m$ member ensemble $X_{v, 1},\ldots, X_{v, m}$. The further variables $w\in \{\text{doy}, \sin, \cos\}$ will be denoted by $X_{w}$. The corresponding realizations are indicated by lowercase letters. The response variable 2\,\si{\meter} surface temperature is represented by $Y_{\mathrm{t2m}}$ with realization $y_{\mathrm{t2m}}$. In the following, we will use the term \textit{reduced variable set} for $\overline{X}_{\mathrm{t2m}}, S_{\mathrm{t2m}}$ including the response variable $Y_{\mathrm{t2m}}$. Furthermore, we designate the set of the mean and standard deviation of the first 10 weather variables in Table \ref{tab: pred_var} incl. the response variable as \textit{extended variable set}. 

\begin{table}[h!]
\begin{center}
\begin{tabular}{c c } 
\toprule
Variable & Description\\ \hline 
t2m & 2\,\si{\meter} surface temperature\\
d2m & 2\,\si{\meter} surface dewpoint temperature\\
pr & surface pressure\\
sr & surface solar radiation\\
u10m & 10\,\si{\meter} surface $u$-wind speed component\\
v10m & 10\,\si{\meter} surface $v$-wind speed component\\
r2m & 2\,\si{\meter} surface relative humidity\\
tcc & total cloud cover\\
ws10m & 10\,\si{\meter} surface wind speed\\
wg10m & 10\,\si{\meter} surface wind gust\\
doy & day of the year \\
sin & sine-transformed day of the year \\
cos & cosine-transformed day of the year \\
\bottomrule
\end{tabular}
\end{center}
\caption{Potential predictor variables.}
\label{tab: pred_var}
\end{table}

Eventually, we will use the period of 2015-2019 as training set and the complete year 2020 as independent validation set. The implementation of some of the methods requires the tuning of specific hyperparameters and selection of marginal distributions for which the final specifications can be found on \href{https://github.com/jobstdavid/paper_gamvinereg}{GitHub}. To avoid overfitting in the model selection process, we further split the training set into the period of 2015-2018 which is used for pure training of the models, while the year 2019 is used for testing in the model selection process. After finalizing the choice of the most suitable model variant based on the testing period, the entire training period from 2015-2019 is used to fit that model for the final evaluation on the validation set. All computations on the data set \parencite{Jobst2023c} will be carried out using the statistical software \texttt{R} running version 3.6.3 by \textcite{RCT2020}.

\section{Ensemble postprocessing methods}
\label{sec: Methods}

In this section, we briefly describe the compared ensemble postprocessing techniques. All methods will be applied locally, i.e. for each station a separate model is estimated.

\subsection{Ensemble model output statistics}
\label{sec: EMOS}
Ensemble model output statistics (EMOS), also known as non-homogeneous regression, is a parametric postprocessing method proposed by \textcite{Gneiting2005}. This approach is based on the idea of distributional regression, assuming a predictive distribution family 
$\mathcal{D}(\mu, \sigma, \nu, \varphi)$, where the parameters $\mu$, $\sigma$, $\nu$, and $\varphi$ indicate location, scale, shape, and degrees of freedom, respectively. There are link functions to connect  parameters with corresponding predictors $\bm{x}_\mu,\bm{x}_\sigma, \bm{x}_\nu, \bm{x}_\varphi$ via $\mu:=h_\mu(\bm{x}_\mu),\sigma:=h_\sigma(\bm{x}_\sigma), \nu:=h_\nu(\bm{x}_\nu), \varphi:=h_\varphi(\bm{x}_\varphi)$ in order to retain  parameter ranges. In this context the predictors 
are typically ensemble members and summary statistics thereof. The predictive distribution is selected with regard to the type of weather quantity, e.g. a Gaussian normal distribution for the variable 2\,\si{\meter} surface temperature as suggested by \textcite{Gneiting2005}. As the logistic and skewed logistic distribution as well as the skew normal distribution show only minor differences with respect to the performance of the Gaussian normal distribution \parencites{Gebetsberger2019, Taillardat2021} we assume the latter in the following, i.e. $Y_{\mathrm{t2m}}\sim \mathcal{N}(\mu, \sigma)$ with location parameter $\mu\in \R$, scale parameter $\sigma>0$ as well as inverse link functions $h_\mu^{-1}:=\text{id}$, $h_\sigma^{-1}:=\log$.

In its basic formulation EMOS uses the reduced variable set with predictors $\overline{X}_{\mathrm{t2m}}$ and $S_{\mathrm{t2m}}$ and  connects the Gaussian (transformed) distribution parameters to the predictors via the linear relationships 

\begin{align}
\mu:=a_0+a_1\overline{x}_{\text{t2m}},\quad \log(\sigma):=b_0+b_1\log(s_{\mathrm{t2m}}).
\label{eq: emos_t2m}
\end{align}
The coefficients $a_0,a_1,b_0,b_1\in \R$ are estimated e.g. by a sliding training window. However, to take the strong seasonal periodic patterns of $Y_{\mathrm{t2m}}$ (e.g. higher values in the summer period, lower values in the winter period) into account and to facilitate a fair comparison with the other methods, we add the sine- and cosine-transformed day of the year $X_{\sin}, X_{\cos}$ as further predictors to the equation of both parameters. Therefore, we assume the conditional predictive distribution 
\begin{align}
f(y_{\mathrm{t2m}}\vert x_{\text{t2m}, 1},\ldots,& x_{\text{t2m}, m}, x_{\sin}, x_{\cos})\sim\mathcal{N}(\mu, \sigma),\\
\mu:=a_0+a_1x_{\sin}+a_2x_{\cos}+a_3\overline{x}_{\text{t2m}},&\quad \log(\sigma):=b_0+b_1x_{\sin}+b_2x_{\cos}+b_3s_{\mathrm{t2m}},
\label{eq: EMOS}
\end{align}
with coefficients $a_i, b_i\in \R$ for $i=0,\ldots, 3$, as e.g. \textcite{Hemri2014, Dabernig2017}. Consequently, we incorporate the seasonality by seasonal varying intercepts as introduced in Equation \eqref{eq: EMOS} in comparison to the basic formulation in Equation \eqref{eq: emos_t2m}.

The coefficients of the parameters in Equation \eqref{eq: EMOS} are estimated by optimizing the sum of a proper verification score over the training period between 2015 and 2018 with the Broyden-Fletcher-Goldfarb-Shannon (BFGS) algorithm. We investigate the optimization with respect to two different scores, namely the CRPS (continuous ranked probability score) and the LogS (logarithmic score). Then, we choose the best performing specification (CRPS or LogS) according to the mean CRPS over all testing days in 2019 and all stations. The implementation is based on the \texttt{R}-package \texttt{crch} by \textcite{Messner2016}.

\subsection{Gradient boosted ensemble model output statistics}
\textcite{Messner2017} suggested an extension of EMOS which allows to select the most relevant predictor variables $X_1,\ldots, X_p$ for the model by a gradient-boosting approach (EMOS-GB). This approach is especially useful if the amount of potential predictor variables is large. Similar to the EMOS model the conditional normal distribution $f(y\vert x_1, ..., x_p)\sim \mathcal{N}(\mu, \sigma)$ with
\begin{align}
    \mu&:=a_0+a_1x_1+\ldots +a_px_p,\quad a_0, a_1,\ldots, a_p\in \mathbb{R},\\
    \log(\sigma)&:=b_0+b_1x_1+\ldots+b_px_p,\quad b_0, b_1, \ldots, b_p\in \mathbb{R},
\end{align}
is assumed. 
We include the  sine- and cosine transformed day of the year $X_{\sin}, X_{\cos}$ into the extended variable set to account for seasonality. This results in $p=22$ predictor variables (see Table \ref{tab: pred_var}) for each distribution parameter. 

The boosting procedure initializes all coefficients for $\mu$ and $\sigma$ at zero and to iteratively updates only the one which corresponds to the predictor improving predictive performance the most. Using the gradient of the loss function, the predictor with the highest correlation to the gradient is selected and then the corresponding coefficient is updated by taking a step in the direction of steepest descent of the gradient. This procedure is carried out until a stopping criterion is reached to avoid overfitting.

 The implementation is based on the \texttt{R}-package \texttt{crch} by \textcite{Messner2016}. We tune the gradient-boosted EMOS (EMOS-GB) model by grid search with respect to the loss function (LogS or CRPS), maximum number of boosting iterations (100, 500, 1000, 2000), stopping criterion (AIC, BIC), and step size (0.05, 0.1, 0.2) on the training data set between 2015 and 2018. Then, we choose the best performing model version according to the mean CRPS over all testing days in 2019 and stations for the validation period. 

\subsection{D-vine copula based quantile regression}

We apply the D-vine copula based quantile regression (DVQR) as proposed by \textcite{Kraus2017} and further described in Section \ref{sec: D-vine GAM copula based quantile regression}, where we use the reduced variable set by minimizing the BIC-corrected conditional log-likelihood.

To take account of the seasonality of our variables, DVQR is first estimated on a refined rolling training period \parencite{Moeller2018} with window size $n\in \mng{10, 15, 20, \ldots, 100}$, for which we use the days $\{t-n, \ldots, t-2, t-1\}$ in the year where the forecast day $t$ lives and the days $\{t-n,\ldots, t-2, t-1, t, t+1, t+2, \ldots, t+n\}$ in the previous $k=4$ years. The optimal window size $n$ for the validation period is determined based on the minimal mean CRPS over all testing days in 2019 and all stations. Afterwards, DVQR is estimated using $k=5$ years with the estimated optimal length $n$. Due to very high computational costs of approximately 7 hours to estimate the models in 2019 for one station and our limited resources of one CPU with 40 cores and 62.5 GB RAM to determine the optimal window length $n$ we can not investigate DVQR on the extended variable set. This high computational burden underlines again the need of an alternative approach such as GAM-DVQR in a higher-dimensional context.

\paragraph{Marginal distributions.} The marginal distributions are fitted via kernel density estimates using the Gaussian kernel.

\begin{figure}[h!]
     \centering
         \includegraphics[scale = 0.5]{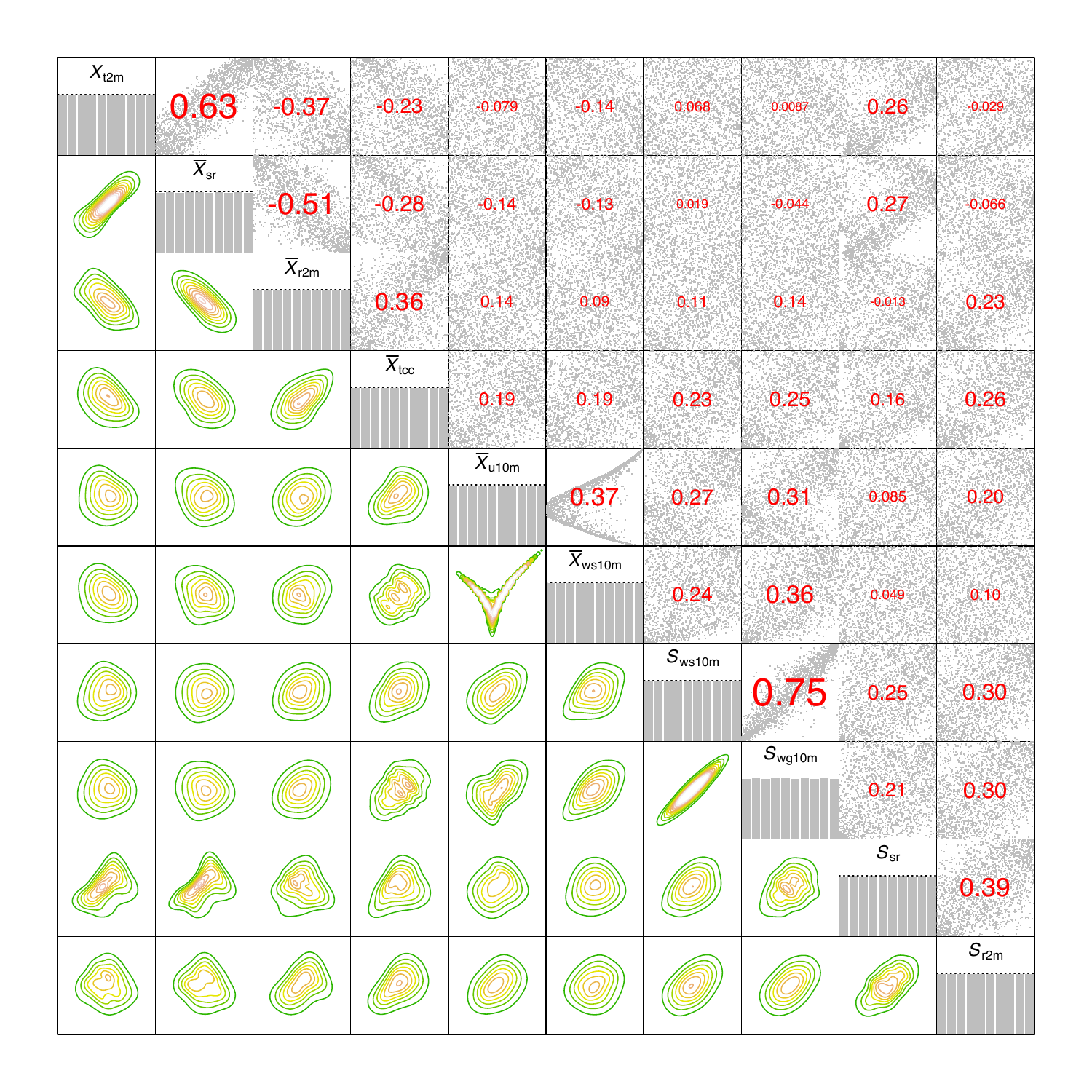}
        \caption{Empirical normalized copula contour plots (lower triangle), PIT histograms (diagonal) and scatterplots including Kendall's $\tau$ correlation (upper triangle) for station Munich in the training data.}
        \label{fig: encp_muc}
\end{figure}

\paragraph{Bivariate copulas.} We allow all bivariate copulas in the \texttt{R}-package \texttt{vinereg}, that is, elliptical copulas (Gaussian and Student-$t$), archimedean copulas as well as rotated versions thereof (Clayton, Gumbel, Frank, Joe, BB1, BB6, BB7 and BB8), and the nonparametric Independence and Transformation Kernel copula (TLL). Elliptical copulas have an elliptical shape in the contour plot. In Figure \ref{fig: encp_muc} for example we can detect a Gaussian copula between $S_{\mathrm{r2m}}$ and $S_{\mathrm{wg10m}}$ and a Student-$t$ copula for the variable pair $S_\mathrm{ws10m}$ and $S_{\mathrm{wg10m}}$. The Gaussian copula has no tail dependence, while the Student-$t$ copula only captures symmetric tail dependence. Any departures from elliptical shapes may indicate to include non-Gaussian dependencies. Therefore, archimedean copulas are provided which exhibit a pear or bone shape in the contour plot allowing to detect lower and/or upper tail dependence (except of the Frank copula). 
In Figure \ref{fig: encp_muc} we see a contour shape indicating a Frank copula between $\overline{X}_{\mathrm{t2m}}$ and $\overline{X}_{\mathrm{sr}}$, a Gumbel copula between $\overline{X}_{\mathrm{r2m}}$ and $\overline{X}_{\mathrm{tcc}}$ and a Clayton copula between $\overline{X}_{\mathrm{ws10m}}$ and $S_{\mathrm{wg10m}}$.
The nonparametric Independence copula has a circular shape in the contour plot (see e.g. between $\overline{X}_{\mathrm{r2m}}$ and $S_{\mathrm{ws10m}}$ in Figure \ref{fig: encp_muc}) and the Transformation Kernel copula can approximate any dependence shape (see e.g. between $\overline{X}_{\mathrm{u10m}}$ and $\overline{X}_{\mathrm{ws10m}}$ in Figure \ref{fig: encp_muc}). Consequently, we cover lots of possible dependence patterns by this copula set.\\ 
The implementation of the DVQR is based on the \texttt{R}-package \texttt{vinereg} by \textcite{Nagler2022}. 

\subsection{D-vine GAM copula based quantile regression}

For the GAM-DVQR as explained in Section \ref{sec: D-vine GAM copula based quantile regression} we consider two cases: Estimation of the GAM-DVQR on the reduced variable set and on the extended variable set.

\paragraph{Marginal distributions.} In both variable sets, we determine the marginal distributions for each variable using distributional regression via generalized additive models for location ($\mu$), scale ($\sigma$) and shape ($\nu$) (GAMLSS, \cite{Rigby2005}). Therefore, we assume for all considered weather variables a distribution $\mathcal{D}(\mu, \sigma, \nu, \varphi)$. As all weather variables show a seasonal periodic behavior we model the distribution parameters by using the notation of Section \ref{sec: EMOS} via 
\begin{align}
	h_\mu^{-1}(\mu)&=a_0+a_1x_{\sin}+a_2x_{\cos},\quad && h_\sigma^{-1}(\sigma)=b_0+b_1x_{\sin}+b_2x_{\cos},\\
	h_\nu^{-1}(\nu)&=c_0,\quad && h_\varphi^{-1}(\varphi)=d_0,
\end{align}
with real valued coefficients using the sine- and cosine transformed day of the year $X_{\sin}, X_{\cos}$ as linear covariates. To keep the marginal models simple and to have comparable settings, the parameters $\nu$ and $\varphi$ are assumed to be constant. For each variable we allow a set of potential distribution families, from which the best is chosen with respect to the mean BIC over all stations in the whole training period. Although \texttt{gamvinereg} allows to select the best performing marginal distribution for each station, we decided against this procedure for a fairer comparison of the methods and a more standardized verification. Besides of the actually tested distribution families, e.g. censored or truncated versions could be additionally analyzed. However, initial tests showed only small differences to the considered distribution families. Additionally, \textcite{Kim2007} outlined in a simulation study that misspecified margins are only problematic for copulas if they are severely misspecified. They studied e.g. fitting normal margins when the true margins are exponential.

\begin{figure}[h!]
	\begin{center}
		\includegraphics[scale = 0.6]{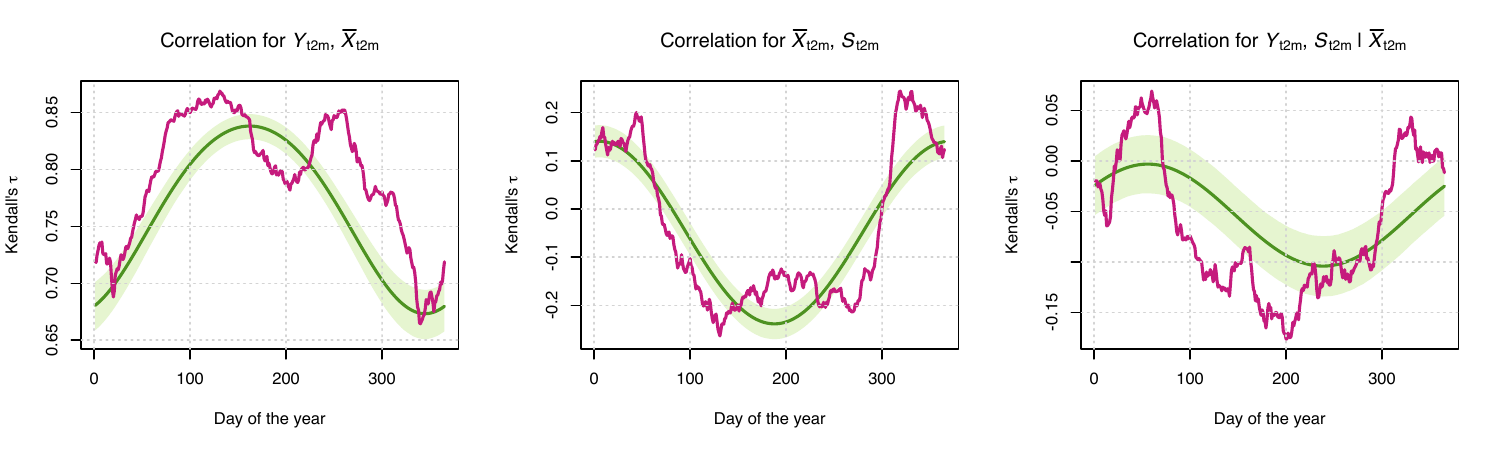}
	\end{center}	
	\caption{Empirical Kendall's $\tau$ (purple) using the refined rolling training period of window size $n=25$, Kendall's $\tau$ prediction (darkgreen) and 95\% confidence band (lightgreen) using the time-dependent correlation model (T1) for station F\"urstenzell  in the training data.}
	\label{fig: red_cor_plot}
\end{figure}

\paragraph{Bivariate copulas.}
The GAM-copula family set for modeling the dependencies consists of the Gaussian, Student-$t$, double Clayton type I-IV and double Gumbel type I-IV copula as implemented in the \texttt{R}-package \texttt{gamCopula} by \textcite{Vatter2020}. The double Clayton and Gumbel copula type consist of additional rotated versions of the Clayton and Gumbel copula, respectively to cover negative dependence as well. All predictor variables and therefore copulas will be selected by minimizing the BIC-corrected conditional log-likelihood as for DVQR. In both, the reduced and extended variable set, each Kendall's $\tau$ linked to a pair-copula is modeled by two different approaches: We assume either a constant or a time-dependent correlation in Equation \eqref{eq: tau_par} and link it to the covariates by one of the following linear models without a non-linear component

\begin{align}
g^{-1}(\tau(\bm{u}, \bm{v}; \bm{\alpha}, \bm{s}))=
\begin{cases}
\alpha_0,& \text{constant correlation (C)},\\
\alpha_0 + \alpha_1 u_{\sin}+\alpha_2 u_{\cos}, & \text{time-dependent correlation (T1)},
\end{cases}
\label{eq: temporal_tau I}
\end{align}
where $\alpha_0, \alpha_1, \alpha_2\in \R$, and covariates $u_{\sin}, u_{\cos}$ denote the sine- and cosine transformed day of the year. The need for a time-adaptive correlation between the predictor variables is illustrated in Figure \ref{fig: red_cor_plot}, where the empirical Kendall's $\tau$ correlation as well as its predictions clearly change over the day of the year. This aspect will be further outlined in Section \ref{sec: Results}.

As the amount of predictor variables can become large in the extended variable set it might not be clear anymore how the temporal correlation can be appropriately modelled by a linear component. Thus, we additionally investigate a time-dependent correlation model using a cyclic cubic spline $s$ depending on the covariate day of the year (doy) via 
\begin{align}
g^{-1}(\tau(\bm{u}, \bm{v}; \bm{\alpha}, \bm{s})))=
s(v_\text{doy}),\quad \text{time-dependent correlation (T2)}.
\label{eq: temporal_tau II}
\end{align}
With this time-dependent non-linear correlation model we can take account of even more flexible (unknown) changes in Kendall's $\tau$ than by a linear model. This approach could be beneficial in higher trees of the D-vine as well. 

Consequently, with GAM-DVQR we introduce an ensemble postprocessing model, which is able to select the most important predictor variables from a large set, taking account of the temporal correlation changes among the predictor variables at the same time. As this model is estimated only once on a static training period, it benefits from a longer consistent training period, while being more efficient than DVQR using a sliding training window.

\begin{table}[h!]
\begin{center}
\resizebox{15cm}{!}{
\begin{tabular}{c c c c} 
\toprule
Model & Marginal specifications & Correlation specifications & Variable  set\\ \hline
GAM-DVQR-C & GAMLSS & constant correlation (C) & reduced \& extended\\
GAM-DVQR-T1 & GAMLSS & time-dependent correlation (T1) & reduced \& extended\\
GAM-DVQR-T2 & GAMLSS & time-dependent correlation (T2) & extended\\
\bottomrule
\end{tabular}
}
\end{center}
\caption{Overview of marginal and correlation parameter specifications.}
\label{tab: overview marginal correlation}
\end{table}

An overview of the selected marginal and correlation parameter specifications can be found in Table \ref{tab: overview marginal correlation}. The estimation of GAM-DVQR is carried out by using the \texttt{R}-package \texttt{gamvinereg} of \textcite{Jobst2023b}.

\section{Verification methods}
\label{sec: Verification}

\textcites{Gneiting2005, Gneiting2007} claim, that the general aim of probabilistic forecasting is to maximize the sharpness of the predictive distribution subject to calibration. \textit{Calibration} refers to the statistical consistency between the predictive cumulative distribution function (CDF) $F$ and the associated observation $Y$. \textit{Sharpness} concerns the spread of the predictive distribution $F$. The more concentrated the forecast, the sharper the forecast, and the sharper the better, subject to calibration. In the following, we present methods to measure calibration and sharpness which are be used in the subsequent application. 

\begin{figure}[h!]
	\begin{center}
		\includegraphics[scale = 0.65]{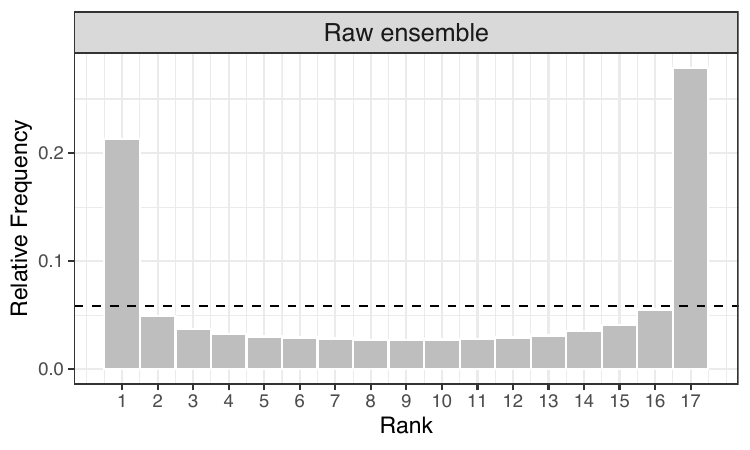}
	\end{center}	
	\caption{Verification rank histogram of the raw ensemble aggregated over all stations and time points in the validation period.}
	\label{fig: vrh_raw}
\end{figure}

\paragraph{Visual assessment of calibration.}

\textcites{Dawid1984, Gneiting2007a} call a continuous predictive probabilistic forecast $F$ calibrated if $F(Y)$ is uniformly distributed.  A so called probability integral transform (PIT) histogram can be used as visual tool for the evaluation of the calibration, where the PIT values are received by evaluating the predictive CDF $F$ at the validating observations. Any departures from uniformity of the PIT histogram can indicate that the predictive distribution $F$ is miscalibrated in some way. A discrete counterpart of the PIT histogram is the so called verification rank histogram displaying the histogram of ranks of observations with respect to the corresponding ordered ensemble forecasts \parencite{Talagrand1997}. In the case of a calibrated $m$-member ensemble, the ranks should be uniformly distributed on the set $\{1,\ldots, m+1\}$.

\paragraph{Uncertainty quantification.}
A further tool for assessing the calibration of a predictive distribution is 
the coverage of a $(1-\alpha)\cdot 100\%$ central prediction interval, $\alpha\in (0,1)$, which is the proportion of validating observations between the lower and upper $\frac{\alpha}{2}$-quantiles of the predictive distribution \parencite{Gneiting2007}. Assuming a calibrated predictive distribution, then $(1-\alpha)\cdot 100\%$ of observations should fall within the range of the central prediction interval.
Sharpness of a predictive distribution can be validated using the width of a $(1-\alpha)\cdot 100\%$ central prediction interval \parencite{Gneiting2007}. Sharper distributions correspond to narrower prediction intervals. Having a $m$-member forecast ensemble, we use a $\frac{m-1}{m+1}\cdot 100\%$ central prediction interval corresponding to the nominal coverage of the raw forecast ensemble and consequently allowing for a direct comparison of all probabilistic forecasts. The target coverage rate for an $m=50$ member ensemble is approximately 96.08\%.

\paragraph{Scoring rules.}
Proper scoring rules rate calibration and sharpness properties simultaneously and thus play important roles in the comparative evaluation and ranking of competing forecasts \parencite{Gneiting2007a}.
An attractive proper scoring rule in weather forecasting is the \textit{continuous ranked probability score} (CRPS, \cite{Matheson1976}), which is defined as 
\begin{align}
\text{CRPS}(F,y):=\int\limits_{-\infty}^{\infty}(F(z)-\mathds{1}\{z\geq y\})^2\, \mathrm{d}z,	\label{eq: CRPS1}
\end{align}
where $F$ is the predictive cumulative distribution function, $y$ is the true/observed value and $\mathds{1}$ denotes the indicator function. \textcite{Gneiting2008} show that, if $F$ has a finite first moment the CRPS can be approximated by
\begin{align}
\text{CRPS}(F,y)\approx \frac{1}{K}\sum\limits_{k=1}^{K}\vert z_k-y \vert -\frac{1}{2K^2}\sum\limits_{k=1}^{K}\sum\limits_{k'=1}^{K}\vert z_k-z_{k'} \vert,	
\end{align}
where $z_k:=F^{-1}\left(\frac{k}{K+1}\right)$, $z_{k'}:=F^{-1}\left(\frac{k'}{K+1}\right)$ for $k,k'\in \{1,\dots, K\}$ and $F^{-1}$ denotes the quantile function of $F$. The mean CRPS over a set of forecast cases is denoted by $\overline{\text{CRPS}}$.

In practice, a probabilistic forecast is sometimes reduced to a point forecast via a statistical summary function such as the mean or median. In this situation, \textit{consistent scoring functions} provide useful tools for forecast evaluation and generate proper scoring rules \parencite{Gneiting2011}. For a set of $n$ forecasts cases we employ 

\begin{equation}
\text{RMSE}:=\sqrt{\frac{1}{n}\sum\limits_{i=1}^{n}(\text{mean}(F_i) - y_i)^2} \quad \text{and}\quad 
\text{MAE}:= \frac{1}{n}\sum\limits_{i=1}^{n}
\vert \text{median}(F_i) - y_i \vert.
\end{equation}

The relative improvement of a forecast with respect to a given reference forecast in terms of CRPS can be quantified by the \textit{continuous ranked probability skill score} (CRPSS) via
\begin{align}
	\text{CRPSS}:=1-\frac{\overline{\text{CRPS}}}{\overline{\text{CRPS}}}_{\text{ref}},
\end{align}
where $\overline{\text{CRPS}}_\text{ref}$ denotes the $\overline{\text{CRPS}}$ of the reference forecast.

\paragraph{Statistical tests to compare predictive performance.}
To evaluate the statistical significance of the differences in the forecasts between two competing postprocessing models, we make use of the \textit{Diebold-Mariano test} \parencite{Diebold1995} for the verification score time series of both models separately at each station. Afterwards we use the \textit{Benjamini-Hochberg procedure} \textcite{Benjamini1995} suggested by \textcite{Wilks2016} that allows to account for multiple testing regarding different stations and to control the overall probability of type I error, for which we choose $\alpha = 0.05$ in the subsequent analysis.

For the verification of the methods we use the \texttt{R}-package \texttt{eppverification} by \textcite{Jobst2021}.

\paragraph{Visual dependence assessment.}

For visually assessing the dependence of two variables $Y$ and $X$ a so called \textit{empirical normalized bivariate contour plot} (see, e.g. Figure \ref{fig: encp_muc}) can be used by the \texttt{R}-package \texttt{VineCopula} of \textcite{Nagler2020b}. This plot is obtained by an approximation of the copula density $c$ and to visualize the contours of the bivariate density function
\begin{align}
    d(z_Y, z_X):=c(\Phi(z_Y), \Phi(z_X))\phi(z_Y)\phi(z_X),
\end{align}
for the $\Phi^{-1}$ transformed copula data $Z_Y:=\Phi^{-1}(F_Y(Y))$ and $Z_X:=\Phi^{-1}(F_X(X))$. For more details concerning these plots, see, e.g. \textcite{Czado2019}. 

\section{Results}
\label{sec: Results}

In the following two subsections the results of the considered methods based on the reduced end extended variable set are presented and discussed. Note, that we additionally investigated time series models as marginal distributions for the GAM-DVQR. However, the respective results turned out to be worse than the ones presented here so we do not show them. 

\subsection{Reduced variable set}
\label{sec: Reduced variable set}

\begin{figure}[h!]
	\begin{center}
		\includegraphics[scale = 0.4]{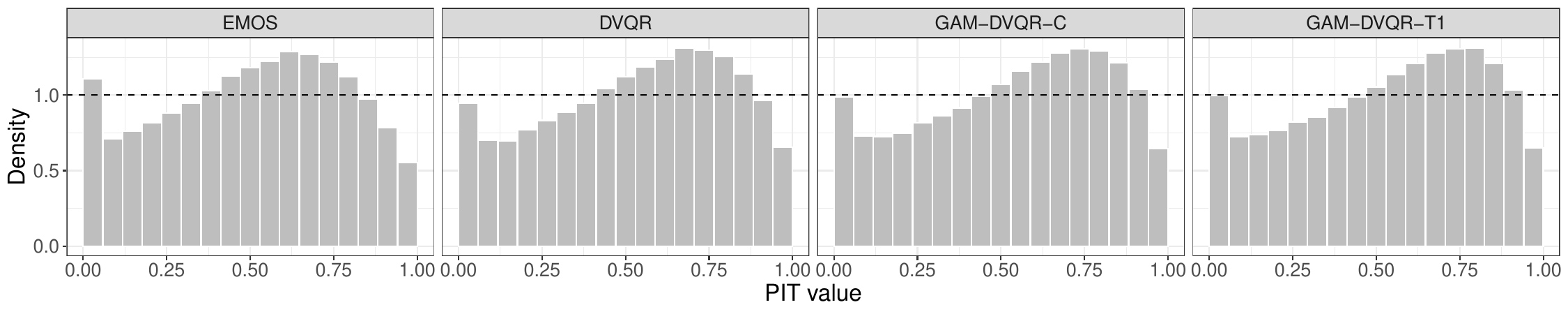}
	\end{center}	
	\caption{PIT histograms of the considered methods, where the PIT values are aggregated over all stations and time points in the validation period.}
	\label{fig: red_pit}
\end{figure}

In this setting, we compare the raw ensemble, EMOS, DVQR and GAM-DVQR on the reduced variable set. Figure \ref{fig: red_pit} shows the PIT histograms for all postprocessing methods. All methods show improved calibration properties in comparison to the raw ensemble in Figure \ref{fig: vrh_raw}. However, they are all slightly skewed to the right causing a small overdispersion more or less in the middle of the PIT histograms and underdispersion at both ends. This impression is supported  by the values of the coverage score shown in Table \ref{tab: red_scores}. 

\paragraph{Verification scores.}

\begin{figure}[h!]
	\begin{center}
		\includegraphics[scale = 0.4]{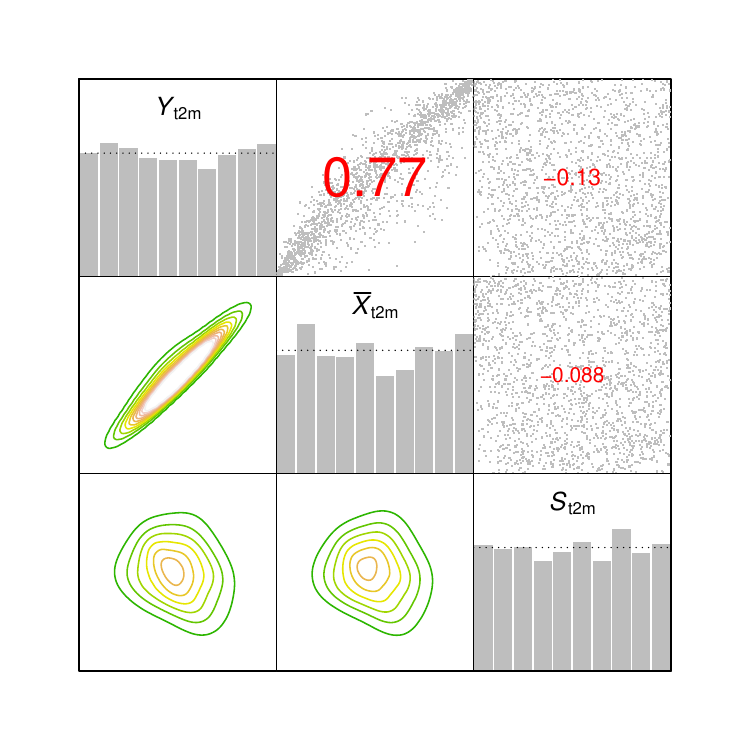}
	\end{center}	
	\caption{Empirical normalized contour plots (lower triangle), PIT histograms (diagonal) and scatterplots including Kendall's $\tau$ correlation (upper triangle) for station F\"urstenzell in the training data.}
	\label{fig: red_encp}
\end{figure}

When looking at Table \ref{tab: red_scores}, we observe that all methods improve the raw ensemble with respect to CRPS by around 20\%-29\%, MAE around 12\%-22\% and RMSE around 13\%-22\%. While EMOS yields the lowest MAE and width, GAM-DVQR-T1 yields the lowest CRPS and RMSE. 
Furthermore, using the procedure for testing the significant differences in the performance between two methods (see Section \ref{sec: Verification}), GAM-DVQR-T1 significantly outperforms EMOS at around 8\% of all stations with respect to CRPS. A reason for the better performance of GAM-DVQR-T1 over EMOS might be that the GAM-DVQR models use non-Gaussian copulas (see Figure \ref{fig: red_encp}), while the basic assumption of EMOS is the Gaussian dependence. The dependence between $Y_{\mathrm{t2m}}$ and $\overline{X}_{\mathrm{t2m}}$ in Figure \ref{fig: red_encp} can be described by a Student-$t$ copula, while for the relationship between $\overline{X}_{\mathrm{t2m}}$ and $S_{\mathrm{t2m}}$ a Gumbel-270$^\circ$ (e.g. double Gumbel type II) or Clayton-90$^\circ$ (e.g. double Clayton type I) copula could be estimated.  A further reason for improved performance in comparison to DVQR could be traced back to the longer and more consistent training data which might lead to more stable estimations \parencite{Lang2020} for GAM-DVQR-T1, while DVQR is estimated on a sliding window.

\begin{table}[h!]
\begin{center}
\begin{tabular}{c c c c c c} 
\toprule
Method & CRPS & MAE & RMSE & Coverage & Width \\ \hline
Raw ensemble &  1.017 &  1.255 &  1.730 & 63.071 &  2.922 \\ \hline
EMOS & 0.718  & \textbf{0.985}  & 1.387 & 95.573 & \textbf{5.441} \\ 
DVQR & 0.717 & 0.985 & 1.366 & \textbf{96.339} & 5.990 \\  \hline
GAM-DVQR-C & 0.719 & 0.989 & 1.358 & 96.499 & 5.994 \\
GAM-DVQR-T1 & \textbf{0.713} & 0.985 & \textbf{1.350} & 96.594 & 5.889\\
\bottomrule
\end{tabular}
\end{center}
\caption{Verification scores aggregated over all stations and time points in the validation period. Bold values represent the best value for each score.}
\label{tab: red_scores}
\end{table}

\begin{table}[h!]
\begin{center}
\resizebox{15cm}{!}{
\begin{tabular}{c | c c  c | c  c c | c c c} 
\toprule
Model & $\alpha_{1,1,0}$ & $\alpha_{1,1,1}$ & $\alpha_{1,1,2}$ & $\alpha_{1,2,0}$ & $\alpha_{1,2,1}$ & $\alpha_{1,2,2}$ & $\alpha_{2,1,0}$ & $\alpha_{2,1,1}$ & $\alpha_{2,1,2}$ \\ \hline 
GAM-DVQR-T1 & 100 & 12 & 74 & 94 & 16 & 78 & 100 & 29 & 62  \\
\bottomrule
\end{tabular}
}
\end{center}
\caption{Significance of the coefficients for the correlation time-dependent GAM-DVQR models in \% after applying the Benjamini-Hochberg procedure to the $p$-value for each coefficient $\alpha_{i,j,k}$ over all stations. $\alpha_{i,j,k}$ represents the $k$-th coefficient for the $j$-th bivariate copula in the $i$-th tree of the D-vine with respect to Equation \eqref{eq: temporal_tau I}.}
\label{tab: sig. p.val copula}
\end{table}

\paragraph{CRPS comparisons.}

In the following, we compare the methods with respect to CRPS in more detail. The time-dependent correlation GAM-DVQR models (T1) perform better than the GAM-DVQR models with constant correlation (C) in terms of CRPS. These results underline the need of a time-dependent correlation model within the GAM-DVQR framework in this application, and also highlight that GAM-DVQR is able to capture the temporal varying empirical correlation. This is further illustrated in Figure \ref{fig: red_cor_plot} where the empirical and predicted Kendall's $\tau$ for a D-vine GAM copula with order $Y_{\mathrm{t2m}}-\overline{X}_{\mathrm{t2m}}-S_{\mathrm{t2m}}$ are plotted. Moreover, the percentage of the significant coefficients for the correlation Equation \eqref{eq: temporal_tau I} in Table \ref{tab: sig. p.val copula} indicate that the suggested predictors for identifying the changes in the correlation parameter seem appropriate.

\begin{figure}[h!]
	\begin{center}
		\includegraphics[scale = 0.6]{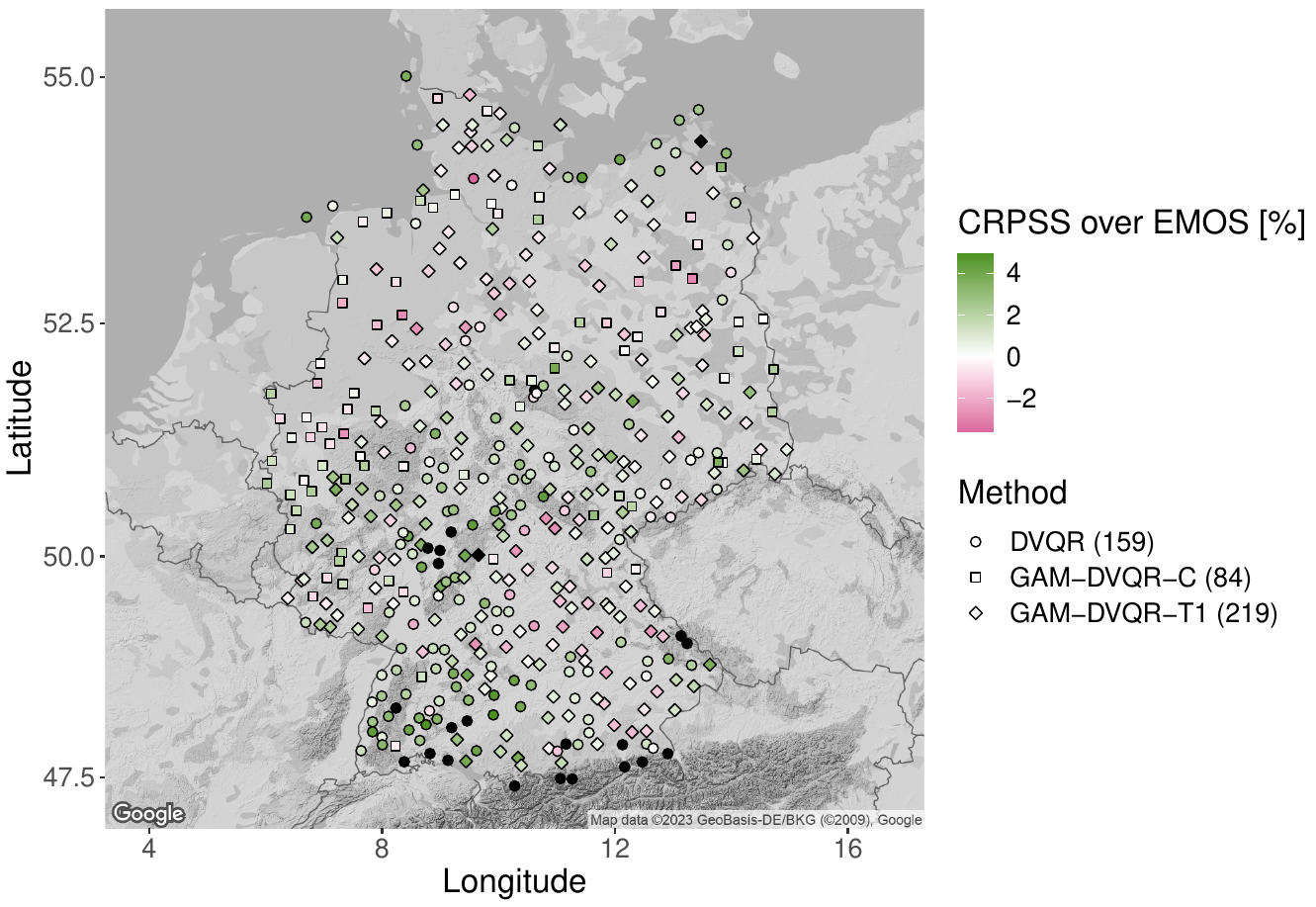}
	\end{center}	
	\caption{Highest CRPSS of the considered methods over EMOS in \% in the validation period. CRPSS $>5\%$ are visualized in black for a better representation. Numbers in brackets denote the count.}
	\label{fig: red_crpss_map}
\end{figure}

We further investigated the station-specific performance of our methods with respect to CRPSS over the benchmark method EMOS. Figure \ref{fig: red_crpss_map} shows the method with the highest CRPSS over EMOS, where the CRPSS values are encoded by colours. The DVQR based methods yield at around 68\% of all stations a positive CRPSS (green colour scale), which implies a substantially better performance of these methods over EMOS. Furthermore, we observe that GAM-DVQR-T1 outperforms the other models most frequently and yields the highest CRPSS at 219 stations, while DVQR is the preferred model at 159 stations.

\subsection{Extended variable set}
\label{sec: Extended variable set}

In this section we investigate the results for the raw ensemble, EMOS, DVQR and GAM-DVQR on the extended variable set.

\begin{figure}[h!]
	\begin{center}
		\includegraphics[scale = 0.4]{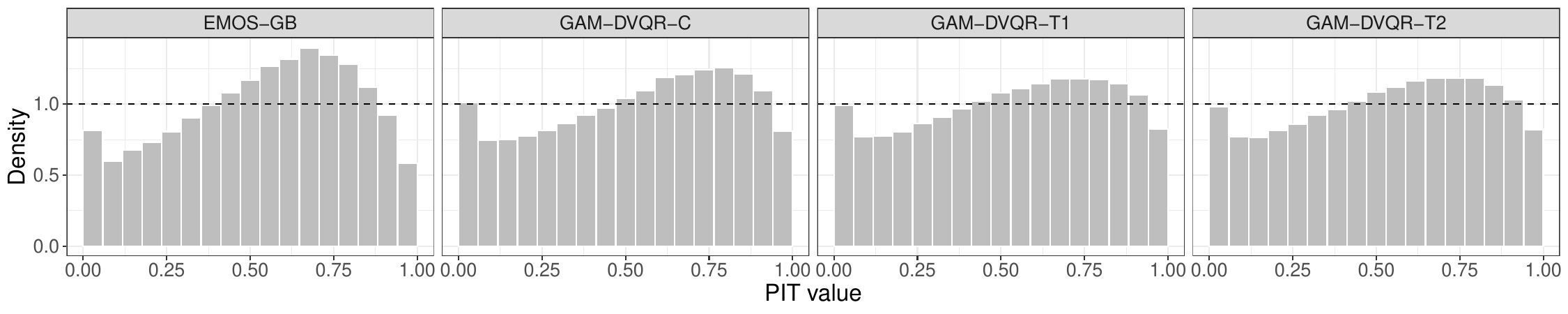}
	\end{center}	
	\caption{PIT histograms of the considered methods, where the PIT values are aggregated over all stations and time points in the validation period.}
	\label{fig: ext_pit}
\end{figure}

The PIT histograms in Figure \ref{fig: ext_pit} indicate that all methods are able to improve the calibration, where the remaining deficiencies are less pronounced than in Section \ref{sec: Reduced variable set}, but still visible. Moreover, GAM-DVQR-T1 and GAM-DVQR-T2 seem to yield the PIT histograms which are closest to a uniform distribution. This impression is further supported by the nearly perfect coverage score of 96.079\% for GAM-DVQR-T1 in Table \ref{tab: ext_scores}. All in all, it appears that the GAM-DVQR models provide a more pronounced calibration in terms of the PIT histograms as well as better coverage values. 

\begin{table}[h!]
\begin{center}
\begin{tabular}{c c c c c c} 
\toprule
Method & CRPS & MAE & RMSE & Coverage & Width \\ \hline
Raw ensemble &  1.017 &  1.255 &  1.730 & 63.071 &  2.922 \\ \hline 
EMOS-GB & 0.706 & 0.979 & 1.336 & 96.756 & 5.670 \\  \hline 
GAM-DVQR-C & 0.710 & 0.980 & 1.337 & 96.145 & 5.680 \\
GAM-DVQR-T1 & 0.684 & 0.942 & 1.282 & \textbf{96.079} & \textbf{5.449}\\
GAM-DVQR-T2 & \textbf{0.681} & \textbf{0.939} & \textbf{1.278} & 96.157 & 5.462 \\
\bottomrule
\end{tabular}
\end{center}
\caption{Verification scores aggregated over all stations and time points in the validation period. Bold values represent the best value for each score.}
\label{tab: ext_scores}
\end{table}

\paragraph{Verification scores.}

All methods are able to improve upon the raw ensemble in terms of CRPS around 30\%-33\%, in terms of MAE around 22\%-25\% and in terms of RMSE around 23\%-26\%, i.e. they all yield a more pronounced improvement in comparison to the methods using only the reduced variable set in Section \ref{sec: Reduced variable set}. This result outlines that an appropriate selection of predictor variables can enhance model performance. Furthermore, it should be pointed out that GAM-DVQR-T2 clearly outperforms all other methods with respect to CRPS, MAE and RMSE followed by GAM-DVQR-T1 for the coverage and width scores in Table \ref{tab: ext_scores}. 

\begin{figure}[h!]
	\begin{center}
		\includegraphics[scale = 0.6]{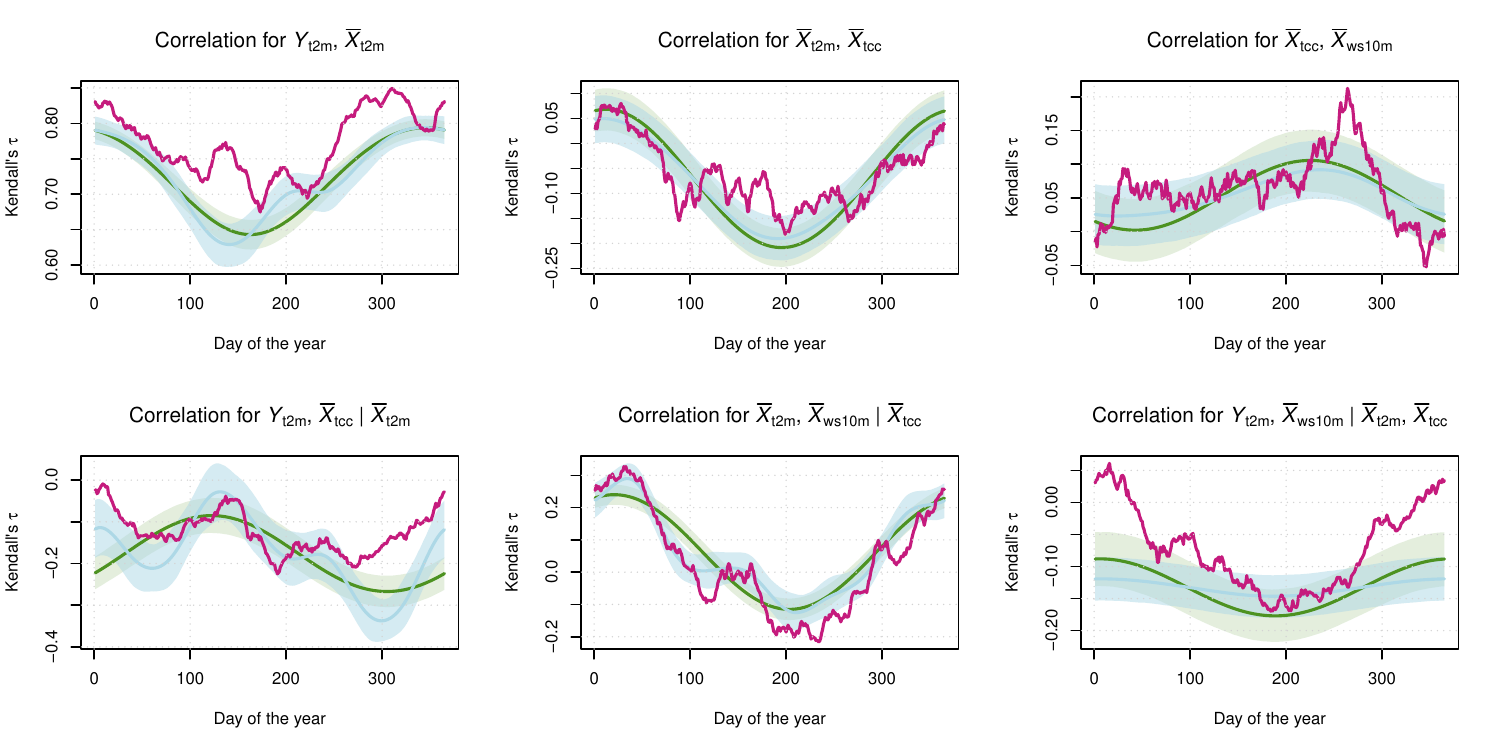}
	\end{center}	
	\caption{Kendall's $\tau$ predictions and its 95\% confidence bands for station Arkona with D-vine GAM copula cutout $Y_{\mathrm{t2m}}-\overline{X}_{\mathrm{t2m}}-\overline{X}_{\mathrm{tcc}}-\overline{X}_{\mathrm{ws10m}}$ in the training data. The green colour represents the linear correlation model (T1), the light blue colour the spline-based correlation model (T2) and the purple color stands for the empirical Kendall's $\tau$ using the refined rolling training period with window size $n=25$.}
	\label{fig: ext_cor_plot}
\end{figure}

We conclude that GAM-DVQR-T1 and GAM-DVQR-T2 perform better than EMOS-GB as well as GAM-DVQR-C, as it can take better account of the temporal variation of the predictor variables and the correlation among them. Based on the results and on Figure \ref{fig: ext_cor_plot} for GAM-DVQR-T2 in comparison to GAM-DVQR-T1, we deduce that the spline based time-dependent correlation model can identify and describe the behavior of the time-varying Kendall's $\tau$ more accurately in a higher-dimensional variable setting than the linear  model for Kendall's $\tau$. Furthermore, if the correlations among the predictor variables are correctly specified, this choice leads to a more reasonable variable selection for GAM-DVQR, and facilitates more appropriate dependencies as well as interactions between the variables than EMOS-GB.

\begin{figure}[h!]
     \centering
     \begin{subfigure}[b]{0.4\textwidth}
         \centering
         \includegraphics[scale = 0.5]{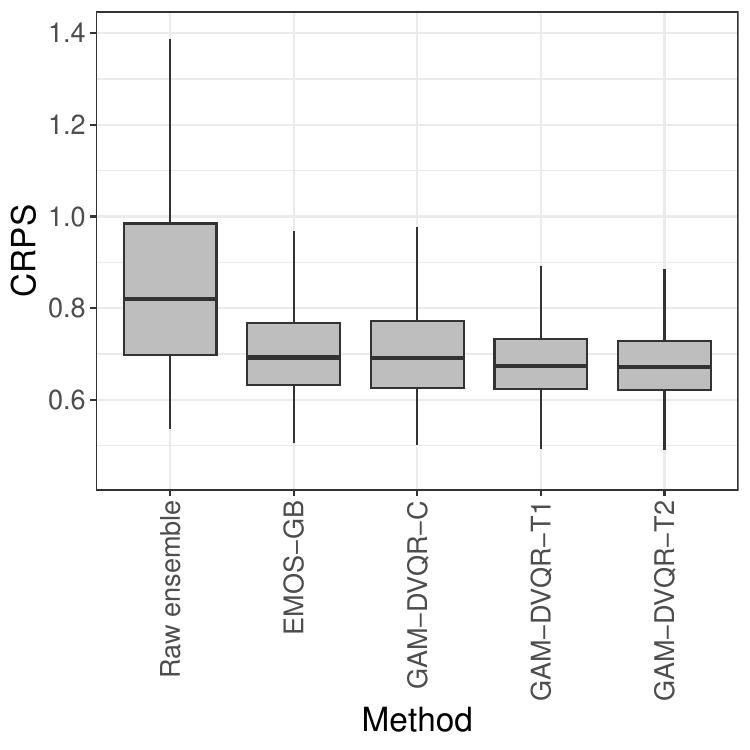}
     \end{subfigure}
		\hspace*{1cm}
     \begin{subfigure}[b]{0.4\textwidth}
         \centering
         \includegraphics[scale = 0.5]{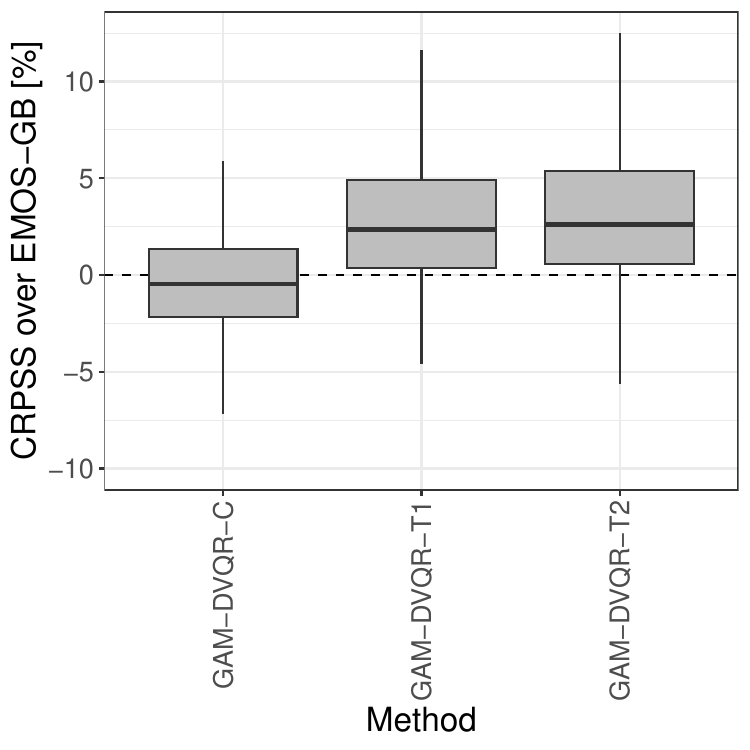}
     \end{subfigure}
        \caption{Left: Boxplots of station-specific mean CRPS values of the considered methods in the validation period. Right: Boxplots of the station-specific CRPSS of the considered methods over EMOS-GB in the validation period. Outliers $\pm 1.5\cdot \mathrm{IQR}$ are omitted for better visual representation.}
        \label{fig: ext_bp_crps}
\end{figure}

\paragraph{CRPS comparisons.}

As in Section \ref{sec: Reduced variable set} we investigate the CRPS in more detail. We observe in the boxplots in Figure \ref{fig: ext_bp_crps} that GAM-DVQR-T1 and  GAM-DVQR-T2 lead to overall lower CRPS values than the other methods. Furthermore, both models have smaller variance in the CRPS values. This point becomes clearer with respect to the CRPSS of the GAM-DVQR methods over EMOS-GB. While GAM-DVQR-C has less variation in the skill scores over EMOS-GB which might be led back to the constant correlation parameter, GAM-DVQR-T1 and GAM-DVQR-T2 show clearly more variance. GAM-DVQR-T2 yields the highest median CRPSS improvement of about 2.7\% over EMOS-GB followed by GAM-DVQR-T1 with about 2.4\% and GAM-DVQR-C with about $-0.5\%$. The results with respect to CRPSS also show, that especially the time-dependent GAM-DVQR models can appropriately capture the time-dependent correlation between the variables. 

\begin{figure}[h!]
	\begin{center}
		\includegraphics[scale = 0.6]{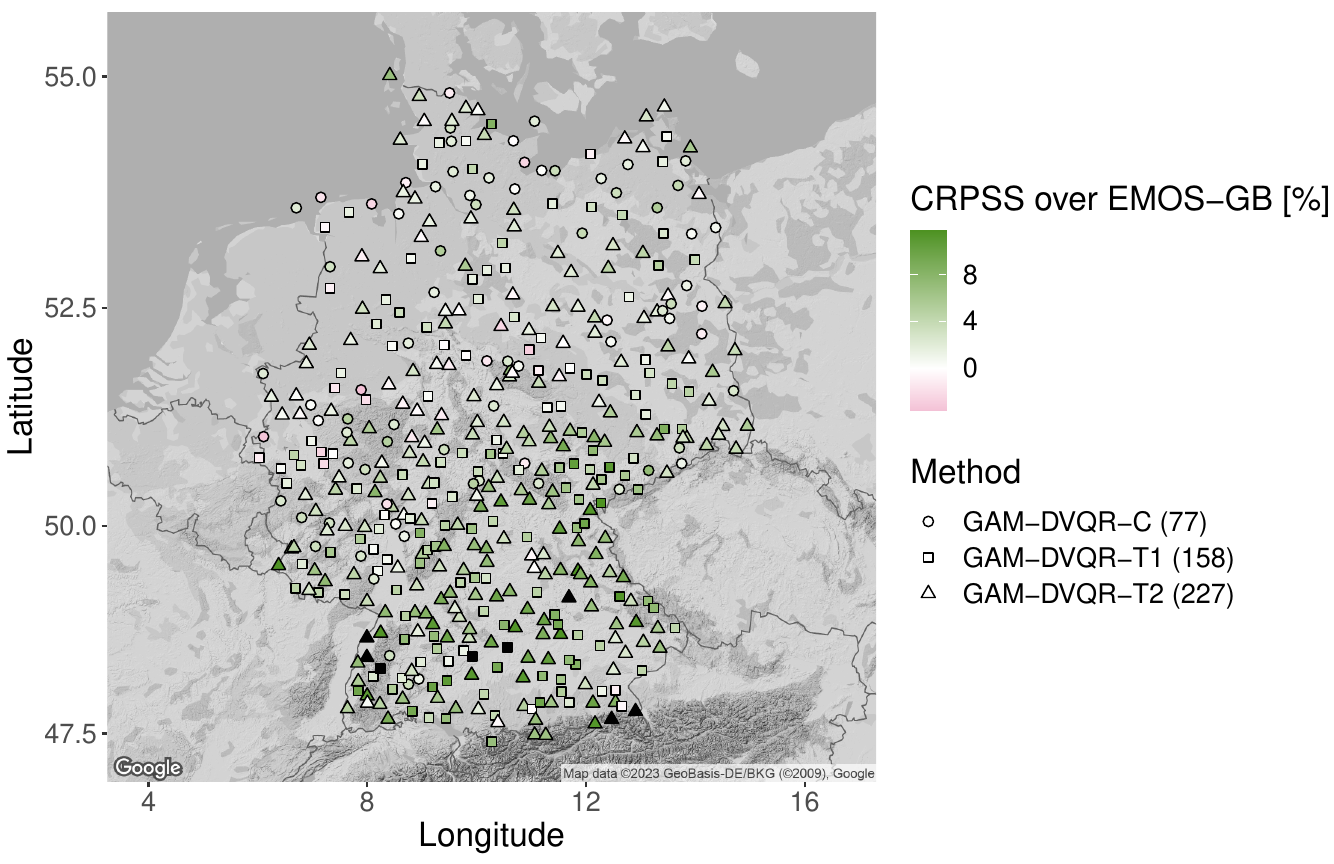}
	\end{center}	
	\caption{Highest CRPSS of the considered methods over EMOS-GB in \% in the validation period. CRPSS $>12\%$ are visualized in black for a better representation. Numbers in brackets denote the count.}
	\label{fig: ext_crpss_map}
\end{figure}

Figure \ref{fig: ext_crpss_map} shows the method with the highest CRPSS in colour over EMOS-GB. It should be pointed out that the CRPSS of the considered methods over EMOS-GB seems to depend on the elevation of the stations. The higher the stations are located, i.e. the more we go into the south of Germany, the higher the CRPSS of the GAM-DVQR methods over EMOS-GB become. At around 89\% of the stations, the GAM-DVQR methods perform better with respect to CRPSS than EMOS-GB. Moreover, GAM-DVQR-T2 outperforms EMOS-GB at around 49\% of all stations in terms of CRPSS, followed by GAM-DVQR-T1 (34\%) and GAM-DVQR-C (17\%).

\begin{figure}[h!]
	\begin{center}
		\includegraphics[scale = 0.5]{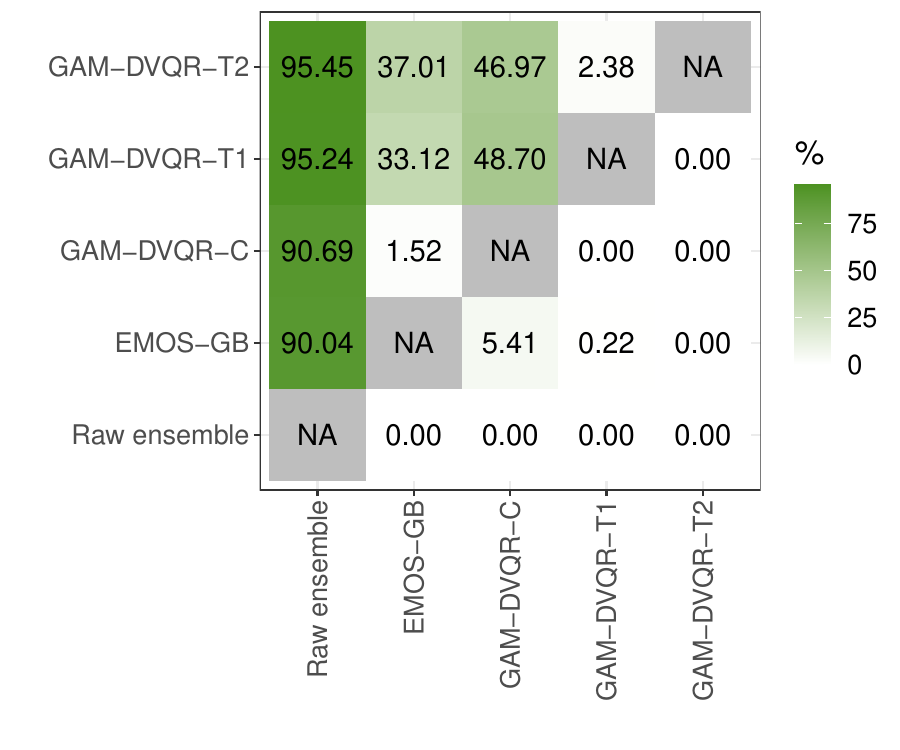}
	\end{center}	
	\caption{Percentage of pair-wise Diebold-Mariano (DM) tests for the 2\,\si{\meter} surface temperature forecasts indicating statistically significant CRPS differences after applying a Benjamini-Hochberg procedure to account for multiple testing for a nominal level of 0.05 of the corresponding one-sided tests.}
	\label{fig: red_sig_table}
\end{figure}

\paragraph{Testing CRPS differences.}

Finally, we have a look at the statistical significance of the differences in the predictive performance with respect to CRPS between the methods. In Table \ref{fig: red_sig_table} the $(i, j)$-entry in the $i$-th row and $j$-th column indicates the percentage of tests where the null hypothesis of equal predictive performance of the corresponding one-sided DM test is rejected in favor of the model in the $i$-th row when compared to the model in the $j$-th column. The remainder of the sum of $(i,j)$- and $(j,i)$-entry to 100\% is the percentage where the score differences are not significant.

All ensemble postprocessing models significantly outperform the raw ensemble for more than 90\% of all stations with respect to CRPS. Furthermore, GAM-DVQR-T2 yields significantly lower CRPS values than EMOS-GB for one third of all stations, followed by GAM-DVQR-T1 (33.12\%) and GAM-DVQR-C (1.52\%). It should also be highlighted that EMOS-GB (90.04\%) performs at around 8\% of the stations significantly worse in comparison to the raw ensemble than EMOS (98.70\%), while GAM-DVQR-T1/T2 (95.24/95.45\%) shows only at 1\% of all stations on the extended variable set significantly lower CRPS values in comparison to its version on the reduced variable set (96.32\%). We conclude that the GAM-DVQR-T1/T2 models lead to more stable results over the raw ensemble compared to EMOS and its gradient-boosted extension EMOS-GB, regardless of whether the reduced or extended variable set is used.

\section{Conclusion and outlook}
\label{sec: Conclusion and outlook}

D-vine GAM copula quantile regression (GAM-DVQR) is a powerful statistical method which allows to select important predictor variables, to model nonlinear relationships between the considered variables and to simultaneously take account of covariate effects linked to the Kendall's $\tau$ of a pair-copula. We complement the presentation of this new method with the \texttt{R}-package \texttt{gamvinereg} by \textcite{Jobst2023b}.

In the application for ensemble postprocessing of 2\,\si{\meter} surface temperature forecasts, GAM-DVQR is able to capture temporal correlation and to choose predictor variables accordingly. Furthermore, the main reasons for the overall better performance of GAM-DVQR over the other methods can be traced back to the modeling of the temporal behavior of the marginal distributions as well as the correlations among the predictor variables. For the reduced and extended variable set, the correlation time-dependent GAM-DVQR models outperform the constant correlation GAM-DVQR models. This  indicates the presence of a non-constant correlation among the variables which needs to be included into the model. Furthermore, the time varying GAM-DVQR yields significant improvements over the benchmark methods EMOS and EMOS-GB. Due to the static training period used for GAM-DVQR in comparison to the conventional day-by-day sliding training window for DVQR, the estimation procedure is more economical, can even result in better fits and makes it appealing for practical and possibly operational use. In future research, we will investigate an extension of the GAM-DVQR method allowing for spatial and spatio-temporal effects for Kendall's $\tau$. This can be specifically relevant in the field of ensemble postprocessing, and the new approach can be compared with other spatial or spatio-temporal postprocessing models, such as e.g. Markovian EMOS by \textcite{Moeller2015}. It might also be beneficial to test other covariate effects besides of the mentioned ones, and to use GAM-DVQR for the postprocessing of non-Gaussian weather quantities, such as wind speed or precipitation. 

In terms of the method itself, the extension of GAM-DVQR to very high-dimensional settings is on the top of our agenda. The work of \textcite{Sahin2022} can serve as a starting point and we plan to compare our extension with suitable methods in various fields. 

Additionally, the method could be further refined to deal with discrete variables, where the work of \textcite{Panagiotelis2012} can be considered. Last but not least, the extension of our method to more general vine structures, e.g. C-vine (canonical vine) or R-vine (regular vine) based on \textcites{Tepegjozova2022, Zhu2021}, respectively, would allow for different dependence structures.

\section*{Acknowledgements}
We are grateful to the European Centre for Medium-Range Weather Forecasts (ECMWF) and the German Weather Service (DWD) for providing forecasts and observation data, respectively. Furthermore, the authors acknowledge support of the research by Deutsche Forschungsgemeinschaft (DFG) Grant Number MO 3394/1-1, and by the Hungarian National Research, Development and Innovation Office under Grant Number NN125679. Annette M\"oller acknowledges support by the Helmholtz Association's pilot project ``Uncertainty Quantification''.

\newpage
\appendix
\begin{Large}
\textbf{Appendix}	
\end{Large}

\section{Hyperparameter specifications}
\label{app: Hyperparameter specifications}

\begin{table}[h!]
\begin{center}
\begin{tabular}{l l l} 
\toprule
\textbf{Method} & Hyperparameter & Value\\ \hline 
\textbf{EMOS} & loss function & CRPS \\
& number of max. iterations & 5000 \\
& stopping criterion & relative threshold $1\mathrm{e}^{-8}$\\ \hline 
\textbf{EMOS-GB} & loss function & LogS \\
& number of max. iterations & 500 \\
& step size & 0.05 \\
& stopping criterion & AIC\\ \hline 
\textbf{DVQR} & window size $n$ & 25 \\
\bottomrule
\end{tabular}
\end{center}
\caption{Overview of the hyperparameter specifications.}
\label{tab: overview_config}
\end{table}

\newpage
\section{Marginal distributions}
\label{app: Marginal distributions}

\flushbottom

Distribution $\mathcal{D}$-sets:\\
\begin{itemize}
	\item $A:=\mng{\mathcal{N}(\mu, \sigma), \mathcal{S}\mathcal{N}(\mu, \sigma, \nu), \mathcal{S}t(\mu, \sigma, \nu, \tau)}$,
	\item $B:=\mng{\text{logit}\mathcal{N}(\mu, \sigma), \text{logit}\mathcal{S}\mathcal{N}(\mu, \sigma, \nu), \text{logit}\mathcal{S}t(\mu, \sigma, \nu, \tau), \mathcal{B}(\mu, \sigma, \nu, \tau)}$,
	\item $C:=\mng{\log\mathcal{N}(\mu, \sigma), \log\mathcal{S}\mathcal{N}(\mu, \sigma, \nu), \log\mathcal{S}t(\mu, \sigma, \nu, \tau)}$,
\end{itemize}
where $\mathcal{N}$ denotes the Gaussian normal distribution, $t$ denotes the Student-$t$ distribution and $\mathcal{B}$ represents the Beta distribution. Furthermore $\mathcal{S}$  abbreviates the skewed version of a distribution, e.g. $\mathcal{SN}$  denotes the skew Gaussian normal distribution and logit as well as $\log$ denote the transformation of the response with the logit- or log-function.

\begin{table}[h!]
\begin{center}

\begin{tabular}{c c c  m{6cm}} 
\toprule
Variable & Tested $\mathcal{D}$-set & Selected $\mathcal{D}$  & Remark\\ \hline
$Y_{\mathrm{t2m}}$ & $A$ & $\mathcal{N}(\mu, \sigma)$ &  \\ \hline
$\overline{X}_{\mathrm{t2m}}$ & $A$ & $\mathcal{N}(\mu, \sigma)$ &   \\
$\overline{X}_{\mathrm{d2m}}$ & $A$ & $\mathcal{S}t(\mu, \sigma, \nu, \tau)$ &  \\
$\overline{X}_{\mathrm{pr}}$ & $A$ & $\mathcal{N}(\mu, \sigma)$ &  \\
$\overline{X}_{\mathrm{sr}}$ & $A$ & $\mathcal{S}\mathcal{N}(\mu, \sigma, \nu)$ &  \\
$\overline{X}_{\mathrm{u10m}}$ & $A$ & $\mathcal{N}(\mu, \sigma)$ &  \\
$\overline{X}_{\mathrm{v10m}}$ & $A$ & $\mathcal{N}(\mu, \sigma)$ &  \\
$\overline{X}_{\mathrm{r2m}}$ & $B$ & $\text{logit}\mathcal{S}t(\mu, \sigma, \nu, \tau)$ &  \\
$\overline{X}_{\mathrm{tcc}}$ & $B$ & $\mathcal{B}(\mu, \sigma, \nu, \tau)$ & raw data transformation based on \textcite{Smithson2006}\\
$\overline{X}_{\mathrm{ws10m}}$ & $C$ & $\log\mathcal{S}t(\mu, \sigma, \nu, \tau)$ &   \\
$\overline{X}_{\mathrm{wg10m}}$ & $C$ & $\log\mathcal{N}(\mu, \sigma)$ &  \\ \hline
$S_{\mathrm{t2m}}$ & $C$ & $\log\mathcal{N}(\mu, \sigma)$ &    \\
$S_{\mathrm{d2m}}$ & $C$ & $\log\mathcal{N}(\mu, \sigma)$ &   \\
$S_{\mathrm{pr}}$ & $C$ & $\log\mathcal{S}t(\mu, \sigma, \nu, \tau)$ &    \\
$S_{\mathrm{sr}}$ & $C$ & $\log\mathcal{S}t(\mu, \sigma, \nu, \tau)$ & \\
$S_{\mathrm{u10m}}$ & $C$ & $\log\mathcal{N}(\mu, \sigma)$ &  \\
$S_{\mathrm{v10m}}$ & $C$ & $\log\mathcal{N}(\mu, \sigma)$ &   \\
$S_{\mathrm{r2m}}$ & $B$ & $\text{logit}\mathcal{N}(\mu, \sigma)$ &  \\
$S_{\mathrm{tcc}}$ & $B$ & $\text{logit}\mathcal{S}t(\mu, \sigma, \nu, \tau)$ & raw data min-max-transformation \& transformation based on \textcite{Smithson2006} \\
$S_{\mathrm{ws10m}}$ & $C$ & $\log\mathcal{N}(\mu, \sigma)$ &   \\
$S_{\mathrm{wg10m}}$ & $C$ & $\log\mathcal{N}(\mu, \sigma)$ &   \\
\bottomrule
\end{tabular}

\end{center}
\caption{GAMLSS distribution selection.}
\end{table}



\newpage
\addcontentsline{toc}{section}{References}
\thispagestyle{plain}
\clearpage

\printbibliography


\end{document}